\documentclass[twocolumn,journal]{IEEEtran}
\usepackage[T1]{fontenc}
\usepackage[latin9]{inputenc}
\usepackage{color}
\usepackage{array}
\usepackage{float}
\usepackage{multirow}
\usepackage{amsmath}
\usepackage{amsthm}
\usepackage{amssymb}
\usepackage{graphicx}
\usepackage[unicode=true,
 bookmarks=true,bookmarksnumbered=true,bookmarksopen=true,bookmarksopenlevel=1,
 breaklinks=false,pdfborder={0 0 0},pdfborderstyle={},backref=false,colorlinks=false]
 {hyperref}
\hypersetup{pdftitle={Your Title},
 pdfauthor={Your Name},
 pdfpagelayout=OneColumn, pdfnewwindow=true, pdfstartview=XYZ, plainpages=false}

\makeatletter

\providecommand{\tabularnewline}{\\}
\floatstyle{ruled}
\newfloat{algorithm}{tbp}{loa}
\providecommand{\algorithmname}{Algorithm}
\floatname{algorithm}{\protect\algorithmname}

\theoremstyle{plain}
\newtheorem{thm}{\protect\theoremname}
\theoremstyle{definition}
\newtheorem{example}[thm]{\protect\examplename}

\usepackage[caption=false,font=footnotesize]{subfig}
\usepackage{algorithmic}
\usepackage{amsmath} 

\DeclareMathOperator*{\argmax}{argmax}

\makeatother

\providecommand{\examplename}{Example}
\providecommand{\theoremname}{Theorem}

\begin{document}

\title{Machine-Learning Beam Tracking and Weight Optimization for mmWave
Multi-UAV Links}

\author{Hsiao-Lan~Chiang, Kwang-Cheng Chen,~\IEEEmembership{Fellow,~IEEE},
Wolfgang Rave, Mostafa Khalili Marandi, and Gerhard Fettweis,~\IEEEmembership{Fellow,~IEEE}\thanks{Hsiao-Lan Chiang and Kwang-Cheng Chen are with the University of South
Florida, Tampa, FL, USA.}\thanks{Wolfgang Rave, Mostafa Khalili Marandi, and Gerhard Fettweis are with
the Vodafone Chair Mobile Communications Systems, Technische Universität
Dresden, Dresden, Germany.}}
\maketitle
\begin{abstract}
Millimeter-wave (mmWave) hybrid analog-digital beamforming is a promising
approach to satisfy the low-latency constraint in multiple unmanned
aerial vehicles (UAVs) systems, which serve as network infrastructure
for flexible deployment. However, in highly dynamic multi-UAV environments,
analog beam tracking becomes a critical challenge. The overhead of
additional pilot transmission at the price of spectral efficiency
is shown necessary to achieve high resilience in operation. An efficient
method to deal with high dynamics of UAVs applies machine learning,
particularly Q-learning, to analog beam tracking. The proposed Q-learning-based
beam tracking scheme uses current/past observations to design rewards
from environments to facilitate prediction, which significantly increases
the efficiency of data transmission and beam switching. Given the
selected analog beams, the goal of digital beamforming is to maximize
the SINR. The received pilot signals are utilized to approximate the
desired signal and interference power, which yield the SINR measurements
as well as the optimal digital weights. Since the selected analog
beams based on the received power do not guarantee the hybrid beamforming
achieving the maximization SINR, we therefore reserve additional analog
beams as candidates during the beam tracking. The combination of analog
beams with their digital weights achieving the maximum SINR consequently
provides the optimal solution to the hybrid beamforming.
\end{abstract}


\section{Introduction}

\begin{figure}[t]
\centering{}\includegraphics[scale=0.38]{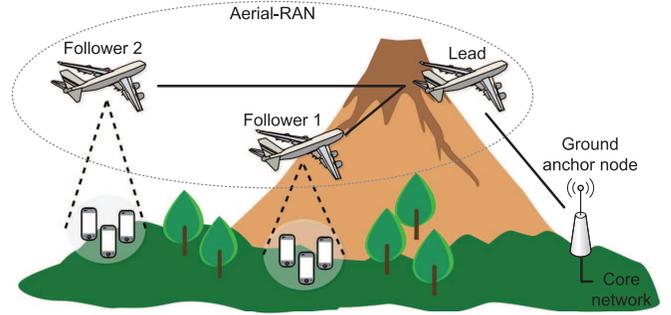}\caption{An example of multi-UAV scenarios. The UAVs are deployed in an area
of interest for search and rescue works, where the lead UAV transmits
the users' data collected from the followers to the ground anchor
node.\label{fig:RAN}}
\end{figure}

Applications of unmanned aerial vehicles (UAVs) in civil uses become
popular in recent years. For example, post-disaster use. A UAV is
capable of carrying a network device as an access point that uses
an intelligent reflecting surface with beamforming to \textit{reflect}
incident signals \cite{Nadeem2019}. A group of UAVs forms an aerial
radio access network (aerial-RAN), which serves short-term network
infrastructure as an independent wireless network or a long-term extension
of existing mobile communication networks \cite{HosseinMotlagh2016,Uluturk2019}.
An aerial-RAN can perform tasks such as (i) the UAVs together transmit
or receive signals from different directions to detect weak signals
from victims and (ii) the UAVs separately serve as independent wireless
networks to provide a wide range of services, see Fig. \ref{fig:RAN}.
In this example, two followers collect data from ground users and
then report the information to the lead UAV, which will pass the data
to a remote ground anchor node \cite{Waharte2010,Zeng2016,Fiandrino2019}.
Such an operation is often characterized by low-latency and high-resilience
constraints. The former is defined as the time to get a response to
information sent, while the latter is the ability that provides and
maintains an acceptable link quality of services in highly dynamic
operations.

Millimeter-wave (mmWave) communication is one of the candidates to
satisfy the low-latency requirement due to availability of large chunks
of spectrum in unlicensed mmWave frequency bands \cite{Rappaport2014,Xiao2017a}.
Compared with sub-6 GHz communications, mmWave propagation suffers
from more severe environmental conditions, such as path loss and a
small number of scattering events \cite{Samimi2015,3GPP38900}. In
order to improve the data rates and quality of service, beamforming
technology for large antenna arrays seems to be a promising approach.
At mmWave frequencies, analog beamforming via a passive phased array
is taken into account due to cost and power consumption concerns \cite{Liberti1999,Hajimiri2005,Bjornson2019}.
With more than one analog beamforming vector, linear combinations
of multiple analog beamforming vectors with weights of digital beamformers
as coefficients provide more degrees of freedom for beamforming designs.
Such a beamforming architecture is called hybrid analog-digital beamforming
\cite{Zhang2005,Ayach2014}.

In hybrid beamforming systems, although both analog and digital beamforming
matrices use the same word beamforming, only the former has a specific
geometrical meaning in the sense of transmitting or receiving signals
towards specific directions in the 3-D space using antenna arrays.
In contrast, the digital weights act in the sense of optimum linear
combining, given some cost criterion. According to the functions of
analog and digital beamforming, hybrid beamforming can be viewed as
first converting a MIMO channel matrix (in the spatial domain) into
an effective channel (in the angular domain) using analog beamforming
vectors \cite{Chiang2017_ICASSP,Chiang2018_JSTSP}. Then, one can
further design the weights of the digital beamformers to linearly
combine the analog beamforming vectors based on some optimality criteria.
Clearly, the performance of hybrid beamforming is dominated by the
analog beam search. In highly dynamic UAV environments with speed
up to 100 m/s \cite{Fotouhi2019}, this challenge (or specifically
speaking, analog beam tracking) will be a critical problem.

One of the key performance indicators for dynamic beam tracking could
be network resilience \cite{Zhou2019}. In dynamic environments, the
UAVs may have to switch the analog beams rapidly in order to stably
provide the acceptable link quality. Given codebooks that consist
of candidates for the analog beams, the work in \cite{Kadur2018}
presented a gradient-based algorithm to find a better beam next to
the currently used beam, and in \cite{Booth2019}, the beam tracking
problem is formulated as a multi-armed bandit problem. One can also
use the extended Kalman filter to recursively track the beams based
on the estimated angles of departure and arrival (AoDs/AoAs) \cite{Va2016}.
 In addition, a conventional object tracking method using reinforcement
learning in computer vision \cite{Supancic2017} has attracted attention
and been used in beam tracking \cite{Klautau2018,Chen2018,Ke2019}.
All above-mentioned methods try to find the beam which can achieve
an acceptable link quality. However, implementing beam tracking for
highly dynamic channels needs a large number of observations (that
is, received pilot signals) by sacrificing the spectral efficiency.
When we pursue a high-resilient multi-UAV communication, the transmission
overhead of pilots is another issue. In this paper, we attempt to
strike the balance between the system resilience and efficiency.

To handle the beamforming problem for a time-varying channel, we let
the UAVs learn how to interact with the highly dynamic environment
during the beam tracking using Q-learning \cite{Watkins1992,Sutton2018}.
Q-learning is a model-free reinforcement learning algorithm that uses
experience, current measurements, and rewards from the environments
to solve the prediction problem without knowing a model of the environment.
When applying Q-learning to beam tracking, the crucial problem is
to design the reward function based on the noisy observations. Please
note that the reward function also influences the experience in Q-learning.
Some prior works in \cite{Klautau2018,Li2019} used true values of
the signal to interference plus noise ratio (SINR) or true values
of the received power to define the reward function, which cannot
faithfully show the performance of Q-learning-based beam tracking
in practical cases. In the proposed method, we use the noisy observations
to design the reward function and take current/past observations as
arguments in such a way to reduce the pilot overhead.

In the analog beam tracking, the analog beams are selected according
to the power of observations.\footnote{Precisely, the power of observations determines the rewards from environments
in Q-learning, and then we use the rewards to find favorable beams.} These beams together yield (nearly) the maximum received power. However,
the spatial-domain interference from different UAVs could seriously
degrade the throughput. Essentially, what really matters to multi-UAV
hybrid beamforming is the SINR maximization \cite{Alkhateeb2015a,Sohrabi2016}.
To this end, given the selected analog beams, one can design the corresponding
digital weights to maximize the SINR. To obtain the measurements of
SINR, we use the received coupling coefficients\footnote{A coupling coefficient is a measure of a pair of analog beamforming
vectors selected on both sides of the channel \cite{Chiang2018_JSTSP}.} (associated with the beams assigned to difference UAVs) to approximate
the desired signal and interference power, which facilitates the design
of the digital weights. Moreover, it is worth noting that the analog
beams leading to the maximum received power may not lead to the maximum
SINR \cite{Chiang2018_JSTSP}. We therefore reserve more candidates
for analog beams during the beam tracking. It turns out that the analog
beams have to be determined after linear combinations of analog beamforming
vectors with the digital weights. 

The\textbf{} \textbf{contributions} of the proposed method are summarized
as follows:
\begin{itemize}
\item \textbf{}The proposed method only requires the received coupling
coefficients as observations to implement both the analog beam tracking
and digital weight optimization. Compared with prior works in the
literature which need detailed knowledge, such as channel, we provide
a more feasible solution to connect multiple UAVs with low complexity.
\item \textbf{}We formulate the beam tracking problem using a Q-learning
model and introduce how to use the coupling coefficients to design
the rewards. The proposed method can stably track the beams in highly
dynamic environments.
\item \textbf{}To track the beams in highly dynamic UAV environments, the
burden of pilot transmission is inevitable. The proposed beam tracking
method uses current and past observations to solve the prediction
problem. In such a way, it significantly increases the efficiency
of data transmission and beam switching.
\item \textbf{}The selected analog beams based on the received power do
not ensure that hybrid beamforming achieves the maximization SINR.
We manage to reserve additional analog beams as candidates during
the beam tracking and then determine which combination of analog beams
with their digital weights achieves the maximum SINR. This idea can
be simply implemented given the coupling coefficients.
\end{itemize}

The rest of this paper is organized as follows: Section II describes
the multi-UAV beamforming system and time-varying AoDs/AoAs. Section
III states the objectives and challenges of the hybrid beamforming
problem in highly dynamic environments. To efficiently track the analog
beams with limited number of observations, Q-learning is applied to
the beam tracking problem for one and multiple links presented in
Section IV. Given selected beam pairs, we pursue the corresponding
optimal digital weights and the solution is provided in Section V.
Simulation results are presented in Section VI, and we conclude our
work in Section VII.

We use the following notations throughout this paper.

\begin{tabular}{cl}
$a$ & A scalar.\tabularnewline
$\mathbf{a}$ & A column vector.\tabularnewline
$\mathbf{A}$ & A matrix.\tabularnewline
$\mathcal{A}$ & A set.\tabularnewline
$[\mathbf{a}]_{n}$ & The $n^{\text{th}}$ entry of $\mathbf{a}$.\tabularnewline
$\mathbf{A}^{*}$ & The complex conjugate of $\mathbf{A}$.\tabularnewline
$\mathbf{A}^{H}$ & The Hermitian transpose of $\mathbf{A}$.\tabularnewline
$\mathbf{I}_{N}$ & The $N\times N$ identity matrix.\tabularnewline
\end{tabular}

\section{System Model}

\begin{figure}[t]
\centering{}\vspace*{-0.3cm}\hspace*{-0.3cm}\includegraphics[scale=0.62]{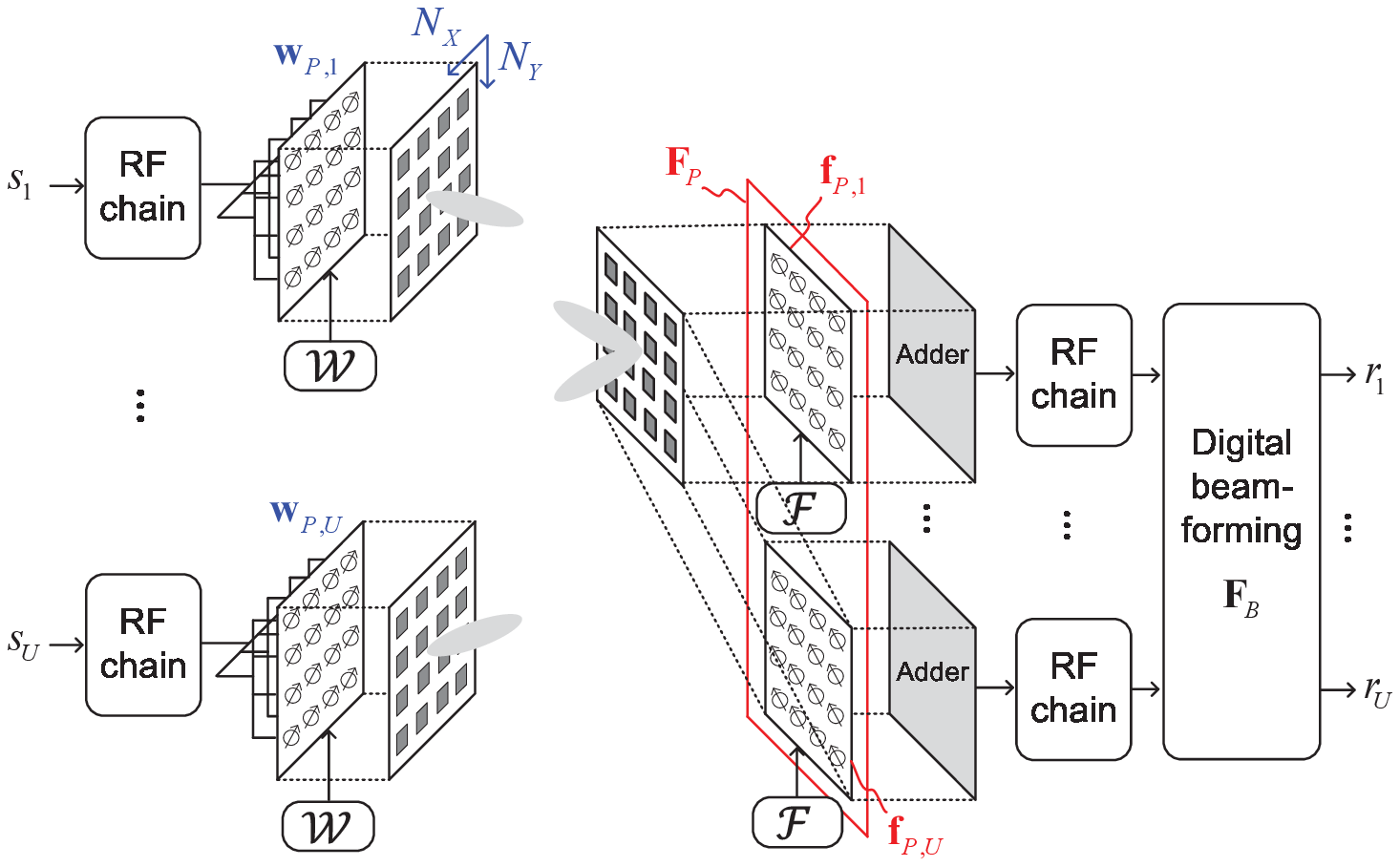}\caption{A multi-UAV hybrid beamforming system has a lead with a hybrid analog-digital
beamformer and $U$ followers equipped with analog beamformers.\label{fig:System-diagram.}}
\end{figure}
A clustered multi-UAV beamforming system shown in Fig. \ref{fig:System-diagram.}
has one lead and $U$ followers. We assume that these UAVs are perfectly
synchronized in time and frequency, and the lead communicates $U$
data streams to $U$ followers at the same time and frequency. That
is, we consider space-division multiple access (SDMA) with beamforming
to enable data transmission/reception for multiple UAVs \cite{Yin2002,Adhikary2014},
and let each UAV be equipped with a uniform rectangular array (URA)
of $N=N_{X}N_{Y}$ antennas.

The goal of multi-UAV beamforming in a highly dynamic environment
is to maximize the system throughput in a discrete time interval  $t=0,\cdots,T$.
At the cluster lead, the signals are received from specific directions
using $U$ analog beamformers at time $t$, denoted by $\mathbf{f}_{P,u,t}\in\mathbb{C}^{N\times1}$,
$u=1,\cdots,U$. The analog beamformers are implemented in the \textit{passband}
as part of the RF front end. Due to the concerns of high implementation
costs and power consumption, they have some limitations, e.g., the
weights of analog beamformers have unit magnitude because analog beamformers
are typically implemented by phase shifters \cite{Hajimiri2005}.
The $U$ analog beamforming vectors together are denoted by the matrix
$\mathbf{F}_{P,t}=[\mathbf{f}_{P,1,t},\cdots,\mathbf{f}_{P,U,t}]\in\mathbb{C}^{N\times U}$,
and these vectors can be further combined with the weights of the
\textit{baseband} digital beamformer $\mathbf{F}_{B,t}\in\mathbb{C}^{U\times U}$. 

Given a pre-defined codebook $\mathcal{F}=\{\tilde{\mathbf{f}}_{n_{f}}\in\mathbb{C}^{N\times1},n_{f}=1,\cdots,N_{F},\:N_{F}>U\}$,
the $U$ analog beamforming vectors at the lead are selected from
the set $\mathcal{F}$. Beam $\tilde{\mathbf{f}}_{n_{f}}$ of the
URA, i.e., the $n_{f}^{\text{th}}$ member of $\mathcal{F}$ can be
represented by the Kronecker product (denoted by $\otimes$) of the
beamforming vectors $\tilde{\mathbf{f}}_{X,n_{f}}\in\mathbb{C}^{N_{X}\times1}$
and $\tilde{\mathbf{f}}_{Y,n_{f}}\in\mathbb{C}^{N_{Y}\times1}$ in
$x$- and $y$-direction respectively \cite{Balanis2005}:
\begin{equation}
\tilde{\mathbf{f}}_{n_{f}}=\tilde{\mathbf{f}}_{X,n_{f}}\otimes\tilde{\mathbf{f}}_{Y,n_{f}},\label{eq: f}
\end{equation}
and the element of $\tilde{\mathbf{f}}_{X,n_{f}}$ and $\tilde{\mathbf{f}}_{Y,n_{f}}$
can be represented by

\begin{figure}[t]
\centering{}\includegraphics[scale=0.38]{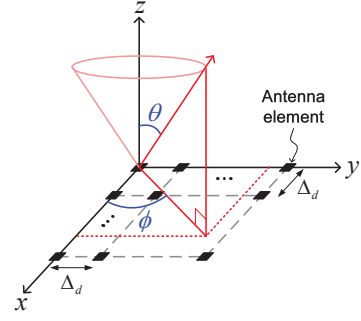}\caption{An array geometry of the URA. \label{fig: URA}}
\end{figure}

{\small{}
\begin{align}
[\tilde{\mathbf{f}}_{X,n_{f}}]_{n_{x}} & =\frac{\exp\left(-j\tfrac{2\pi}{\lambda_{0}}\cos(\phi_{n_{f}})\sin(\theta_{n_{f}})(n_{x}-1)\Delta_{d}\right)}{\sqrt{N_{X}}},\nonumber \\{}
[\tilde{\mathbf{f}}_{Y,n_{f}}]_{n_{y}} & =\frac{\exp\left(-j\tfrac{2\pi}{\lambda_{0}}\sin(\phi_{n_{f}})\sin(\theta_{n_{f}})(n_{y}-1)\Delta_{d}\right)}{\sqrt{N_{Y}}},\label{eq: f_2}
\end{align}
}where $n_{x}=1,\cdots,N_{X}$ and $n_{y}=1,\cdots,N_{Y}$ are the
indices of antenna elements in $x$- and $y$-direction respectively.
Also, $\phi_{n_{f}}$ and $\theta_{n_{f}}$ are respectively the $n_{f}^{\text{th}}$
candidate for the azimuth and elevation steering angles at the lead
(see Fig. \ref{fig: URA}), $\Delta_{d}=\lambda_{0}/2$ is the distance
between neighboring antenna elements, and $\lambda_{0}$ is the wavelength
at the carrier frequency.

For the $U$ followers, each only uses a single analog beamformer
$\mathbf{w}_{P,u,t}\in\mathbb{C}^{N\times1}$ with $N$ phase shifters
to communicate with the lead.\footnote{We assume that all the UAVs are equipped with a hybrid beamforming
architecture since the leading UAV may change over time. The lead
is randomly selected from $U+1$ UAVs at the beginning. } Similar to the analog beams at the lead, each follower selects an
analog beam from codebook $\mathcal{W}=\{\tilde{\mathbf{w}}_{n_{w}}\in\mathbb{C}^{N\times1},n_{w}=1,\cdots,N_{W},N_{W}>U\}$.\footnote{Essentially, these two codebooks are the same, i.e., $\mathcal{W}=\mathcal{F}$.
We specify the beamforming problem in terms of two different notations
of codebooks for generality.}

Via a time-varying channel $\mathbf{H}_{u,t}\in\mathbb{C}^{N\times N}$
between the lead and follower $u$, the received signal at the lead
after the hybrid beamformer is the superposition of the desired signal,
interference from other UAVs, and combined noise \cite{Alkhateeb2015a,Sohrabi2016}:
\begin{flalign}
r_{u,t} & =\underset{\text{desired signal}}{\underbrace{\mathbf{f}_{B,u,t}^{H}\mathbf{F}_{P,t}^{H}\mathbf{H}_{u,t}\mathbf{w}_{P,u,t}s_{u,t}}}\nonumber \\
 & +\underset{\text{interference}}{\underbrace{\mathbf{f}_{B,u,t}^{H}\mathbf{F}_{P,t}^{H}\sum_{i=1,i\neq u}^{U}\mathbf{H}_{i,t}\mathbf{w}_{P,i,t}s_{i,t}}}+\underset{\text{ combined noise}}{\underbrace{\mathbf{f}_{B,u,t}^{H}\mathbf{F}_{P,t}^{H}\mathbf{n}_{t}}},\label{eq: r}
\end{flalign}
where $s_{u,t}\in\mathbb{C}$ is the pilot signal satisfying $|s_{u,t}|^{2}=1$
and $\text{E}[s_{u,t}s_{i,t}^{*}]=0$, $\mathbf{n}_{t}\in\mathbb{C}^{N\times1}$
is an $N$-dimensional circularly symmetric complex Gaussian (CSCG)
random noise vector with mean $\boldsymbol{0}_{N\times1}$ and covariance
matrix $\sigma_{n}^{2}\mathbf{I}_{N}$, i.e., $\mathbf{n}_{t}\sim\mathcal{CN}(\boldsymbol{0}_{N\times1},\sigma_{n}^{2}\mathbf{I}_{N})$,
and $\mathbf{f}_{B,u,t}\in\mathbb{C}^{U\times1}$ is the $u^{\text{th}}$
column of $\mathbf{F}_{B,t}$. 

The link between the lead and follower $u$ is modeled as a line-of-sight
(LoS) path. According to the relative position and orientation between
the transmitter and receiver, the MIMO channel matrix can be determined
by the complex path gain $\rho_{u}\in\mathbb{C}$ and the outer product
of two array response vectors $\mathbf{a}_{A,u,t}\in\mathbb{C}^{N\times1}$
and $\mathbf{a}_{D,u,t}\in\mathbb{C}^{N\times1}$, which are functions
of AoA and AoD \cite{Ayach2014,Sayeed2002}. Thus, the channel matrix
is expressed by
\begin{align}
\mathbf{H}_{u,t} & =\rho_{u}\cdot\mathbf{a}_{A,u,t}\cdot\mathbf{a}_{D,u,t}^{H}.
\end{align}
In a manner similar to the steering vector in (\ref{eq: f}), the
array response vectors can be represented by the Kronecker product
of the array response vectors in $x$- and $y$-direction. Take $\mathbf{a}_{D,u,t}$
as an example:
\begin{equation}
\mathbf{a}_{D,u,t}=\mathbf{a}_{D,X,u,t}\otimes\mathbf{a}_{D,Y,u,t},\label{eq: a}
\end{equation}
and the entries of $\mathbf{a}_{D,X,u,t}$ and $\mathbf{a}_{D,Y,u,t}$
are given by

{\small{}
\begin{align}
[\mathbf{a}_{D,X,u,t}]_{n_{x}} & =\frac{\exp\left(\tfrac{-j2\pi}{\lambda_{0}}\cos(\phi_{D,u,t})\sin(\theta_{D,u,t})(n_{x}-1)\Delta_{d}\right)}{\sqrt{N_{X}}},\nonumber \\{}
[\mathbf{a}_{D,Y,u,t}]_{n_{y}} & =\frac{\exp\left(\tfrac{-j2\pi}{\lambda_{0}}\sin(\phi_{D,u,t})\sin(\theta_{D,u,t})(n_{y}-1)\Delta_{d}\right)}{\sqrt{N_{Y}}},\label{eq: a-1}
\end{align}
}where the random variables $\phi_{D,u,t}$ and $\theta_{D,u,t}$
stand for the azimuth and elevation angles of departure at time $t$.
Given the azimuth and elevation angles of arrival (denoted by $\phi_{A,u,t}$,
$\theta_{A,u,t}$), the array response vector at the receiver (i.e.,
$\mathbf{a}_{A,u,t}$) has a similar form as (\ref{eq: a}).

To model a highly dynamic environment for the angles under an observed
LoS path, a Gaussian random walk is used to generate the time-varying
angles $\phi_{A,u,t}$, $\theta_{A,u,t}$, $\phi_{D,u,t}$, and $\theta_{D,u,t}$.
For instance, the azimuth angle of arrival $\phi_{A,u,t}$ can be
defined by
\begin{equation}
\phi_{A,u,t}=\phi_{A,u,0}+\sum_{i=1}^{t}\lambda_{i},
\end{equation}
where $\phi_{A,u,0}\sim\mathcal{U}(0,2\pi)$ is a randomly selected
initial angle of $\phi_{A,u,t}$ and follows a uniform distribution,
and $\lambda_{i}\sim\mathcal{N}(0,\sigma_{\lambda}^{2})$ is the disturbance
(or white noise) following a normal distribution. The other three
time-varying angles are generated in a similar way.

\section{Problem Statement}

The goal of hybrid beamforming in the multi-UAV system is to maximize
the SINR (or system throughput) during the time interval $[0,T]$.
Meanwhile, after the combiner $\mathbf{F}_{P,t}\mathbf{F}_{B,t}$,
the variance of the combined noise signal is enforced to remain constant,
i.e.,
\begin{flalign}
\text{E}\left[\left(\mathbf{f}_{B,u,t}^{H}\mathbf{F}_{P,t}^{H}\mathbf{n}_{t}\right)\left(\mathbf{f}_{B,u,t}^{H}\mathbf{F}_{P,t}^{H}\mathbf{n}_{t}\right)^{H}\right] & =\sigma_{n}^{2}\;\forall u,t,
\end{flalign}
which leads to a power constraint on the combiner as
\begin{equation}
\mathbf{f}_{B,u,t}^{H}\mathbf{F}_{P,t}^{H}\mathbf{F}_{P,t}\mathbf{f}_{B,u,t}=1\:\forall u,t.\label{eq: constraint}
\end{equation}
Then, by introducing two sets $\mathcal{I_{F}}_{\!,t}$ and $\mathcal{I_{W}}_{\!,t}$
that include promising candidates for the analog beamforming matrices,
we seek $\mathbf{F}_{P,t}$, $\mathbf{F}_{B,t}$, and $\mathbf{W}_{P,t}$
that together achieve the maximum SINR and satisfy the power constraint
from $t=0$ to $t=T$:
\begin{gather}
\sum_{t=0}^{T}\max_{\mathbf{F}_{P,t}\in\mathcal{I_{F}}_{\!,t},\,\mathbf{W}_{P,t}\in\mathcal{I_{W}}_{\!,t}}\:\left\{ \max_{\mathbf{F}_{B,t}}\sum_{u=1}^{U}\;\frac{P_{S,u,t}}{P_{I,u,t}+\sigma_{n}^{2}}\right\} \label{eq: max}\\
\text{s.t. }\mathbf{f}_{B,u,t}^{H}\mathbf{F}_{P,t}^{H}\mathbf{F}_{P,t}\mathbf{f}_{B,u,t}=1\:\forall u,t,\nonumber 
\end{gather}
where $P_{S,u,t}$ and $P_{I,u,t}$ are the power of the desired and
interference signals given by
\begin{align}
P_{S,u,t} & =\left|\mathbf{f}_{B,u,t}^{H}\mathbf{F}_{P,t}^{H}\mathbf{H}_{u,t}\mathbf{w}_{P,u,t}\right|^{2},\\
P_{I,u,t} & =\sum_{i=1,i\neq u}^{U}\left|\mathbf{f}_{B,u,t}^{H}\mathbf{F}_{P,t}^{H}\mathbf{H}_{i,t}\mathbf{w}_{P,i,t}\right|^{2}.
\end{align}

In the paper, we do not assume the channel state information or any
knowledge of AoAs/AoDs is known to the lead. Instead, the\textit{
}required observations are the{\small{} }estimates{\small{} of }\textit{coupling
coefficients} associated with a beam pair $(\tilde{\mathbf{f}}_{n_{f}},\tilde{\mathbf{w}}_{n_{w}})$,
where $\tilde{\mathbf{f}}_{n_{f}}\in\mathcal{F}$ and $\tilde{\mathbf{w}}_{n_{w}}\in\mathcal{W}$.
By correlating the received pilot signals with the known transmitted
ones, we can obtain such observations given by\footnote{The notation of observation $y_{u,t}(n_{f},n_{w})$ is simplified
from its formal expression given by $y_{u,t}(n_{f}=n_{f}(u,t)\in\{1,\cdots,N_{F}\},n_{w}=n_{w}(u,t)\in\{1,\cdots,N_{W}\})$.}
\begin{flalign}
 & y_{u,t}(n_{f},n_{w})\nonumber \\
 & =s_{u,t}^{*}\,\underset{\text{received polit signal}}{\underbrace{\left(\tilde{\mathbf{f}}_{n_{f}}^{H}\mathbf{H}_{u,t}\tilde{\mathbf{w}}_{n_{w}}s_{u,t}+\tilde{\mathbf{f}}_{n_{f}}^{H}\sum_{i=1,i\neq u}^{U}\mathbf{H}_{i,t}\tilde{\mathbf{w}}_{n_{w}}s_{i,t}+\tilde{\mathbf{f}}_{n_{f}}^{H}\mathbf{n}_{t}\right)}}\nonumber \\
 & =\tilde{\mathbf{f}}_{n_{f}}^{H}\mathbf{H}_{u,t}\tilde{\mathbf{w}}_{n_{w}}+\underset{\triangleq z_{t}}{\underbrace{\left(\tilde{\mathbf{f}}_{n_{f}}^{H}\sum_{i=1,i\neq u}^{U}\mathbf{H}_{i,t}\tilde{\mathbf{w}}_{n_{w}}s_{u,t}^{*}s_{i,t}+s_{u,t}^{*}\tilde{\mathbf{f}}_{n_{f}}^{H}\mathbf{n}_{t}\right)}}\nonumber \\
 & =\underset{\text{coupling coefficient}}{\underbrace{\tilde{\mathbf{f}}_{n_{f}}^{H}\mathbf{H}_{u,t}\tilde{\mathbf{w}}_{n_{w}}}}+z_{t},\label{eq: y}
\end{flalign}
where $z_{t}$ denotes the superposition of the combined interference
and noise, and we assume that it follows a complex normal distribution,
i.e., $z_{t}\sim\mathcal{CN}(0,\sigma_{z}^{2})$.

Given the observations $\{y_{u,t}(n_{f},n_{w})\:\forall u,t\}$, the
strategy of solving the problem (\ref{eq: max}) could be, first,
using the observations to find the sets $\mathcal{I_{F}}_{\!,t}$
and $\mathcal{I_{W}}_{\!,t}$ that ideally consist of the optimal
analog beamforming matrices. However, due to the hardware constraint
on the analog beamformer, the beam probing is time-consuming. When
the channel is highly dynamic, the observations acquired early may
become unreliable. How to use the observations to interact with the
highly dynamic environment during the beam probing becomes a crucial
problem. As a result, the idea of \textit{Q-learning} algorithm \cite{Sutton2018}
is borrowed to find appropriate beams (i.e., the members of $\mathcal{I_{F}}_{\!,t}$
and $\mathcal{I_{W}}_{\!,t}$) for time-varying channels. The concept
of Q-learning is to let the UAVs learn the optimal behavior directly
from the interaction with the environment. Once we determine the candidate
sets $\mathcal{I_{F}}_{\!,t}$ and $\mathcal{I_{W}}_{\!,t}$, the
observations associated with the members of $\mathcal{I_{F}}_{\!,t}$
and $\mathcal{I_{W}}_{\!,t}$ are used to generate the corresponding
digital weights and the SINR measurement.

\section{Analog Beam Tracking Using Q-Learning}

\begin{figure}[t]
\begin{centering}
\includegraphics[scale=0.6]{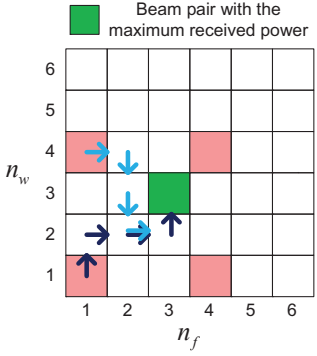}
\par\end{centering}
\caption{All the candidates for the beam pairs $\{(\tilde{\mathbf{f}}_{n_{f}},\tilde{\mathbf{w}}_{n_{w}})\:\forall n_{f},n_{w}\}$
are represented by the grid map, where the red ones are trained during
the initial beam search. An example of the Q-learning-based beam selection
is given in Example \ref{exa:When-starting-from}. According to the
updated Q-values, see Table \ref{tab:The-Q-values}, it will converge
to beam pair $(\tilde{\mathbf{f}}_{3},\tilde{\mathbf{w}}_{3})$ after
few iterations.\label{fig: map}}
\end{figure}
In this section, we introduce an analog beam tracking algorithm for
highly dynamic environments. Starting from a single link between the
lead and a follower, we adopt Q-learning to deal with the beam tracking
problem. The idea can be easily extended to multiple links with additional
constraints.

\subsection{Beam Selection Using Q-Learning for One Link}

To begin with, let us focus on the link between the lead and follower
$u$. That is, we seek the candidates for $\mathbf{f}_{P,u,t}$ and
$\mathbf{w}_{P,u,t}$. When the codebook size is large, the efficient
way of beam tracking is to start from some specific directions that
cover the 3-D environment. This phase is called \textit{initial beam
search}. For example, Fig. \ref{fig: map} shows $N_{F}N_{W}=6\times6$
candidates for the analog beam pair, where $N_{F}$ and $N_{W}$ are
the numbers of elements in codebooks $\mathcal{F}$ and $\mathcal{W}$
respectively. In the example, the four beam pairs highlighted in red
are initially explored. To be formal, we define two sets that consist
of the beams used in the initial search by $\mathcal{F_{I}}=\{\tilde{\mathbf{f}}_{1},\tilde{\mathbf{f}}_{4}\}$
and $\mathcal{W_{I}}=\{\tilde{\mathbf{w}}_{1},\tilde{\mathbf{w}}_{4}\}$
and assume that both the lead and follower have the same initial beam
search pattern. After the beam probing using these four beam pairs,
the one having the maximum received power will be selected as a starting
point of beam tracking in the next phase.

The beam tracking is conventionally implemented by searching a better
choice next to the currently used beam pair \cite{Kadur2018,Yang2019a}.
Both the initial beam search and beam tracking in the above-mentioned
work only explore the environment rather than interact with the environment.
The concept of ``interaction with the environment'' can be viewed
as a beam selection algorithm that can \textit{explore} uncharted
territory and, meanwhile, \textit{exploit} the searching experience.
Concerning a highly dynamic environment, the exploration-exploitation
balance\textit{ }becomes more important to the beam tracking. The
idea of Q-learning is to let an agent (e.g., a UAV) learn to strike
the balance between exploration and exploitation.

In Q-learning, the experience is recorded in a Q-learning table (or
Q-table), see Table \ref{tab:The-Q-values}, which is updated according
to the current measurements. The Q-table is constructed according
to three components: states, actions, and state-action values (also
known as Q-values). Before the learning begins, the state-action values
in the Q-table are initialized to zero. In a state $S_{t}$ at time
$t$, the UAV always implements the following four steps: select an
action $A_{t}$ from the action set $\mathcal{A}=\{\text{up, down, right, left}\}$,
go to the next state $S_{t+1}$, observe a reward $R_{t+1}$, and
update the Q-value, given by \cite[Ch. 6]{Sutton2018}
\begin{equation}
Q(S_{t},A_{t})\leftarrow(1-\alpha)\underset{\text{old value}}{\underbrace{Q(S_{t},A_{t})}}+\alpha\underset{\text{new information}}{\underbrace{\left[R_{t+1}+\gamma\max_{a\in\mathcal{A}}Q(S_{t+1},a)\right]}},\label{eq: Q value}
\end{equation}
where $0<\alpha<1$ is the learning rate (or step size), $0<\gamma<1$
is the discount factor determining the importance of future rewards.
The Q-value update can be described as a weighted average between
the old value and new information.

\begin{table}[t]
\caption{The Q-values are updated according to the states and actions given
in Example \ref{exa:When-starting-from} and Fig. \ref{fig: map}.
Here we let the Q-values be updated by either 0 or 1 for simplicity.\label{tab:The-Q-values} }

\begin{tabular}{|c|c|c|c|c|c|c|c|}
\hline 
\multirow{2}{*}{Time ($t$)} & \multirow{2}{*}{Episode} & Step & State  & \multicolumn{4}{c|}{Action $A_{t}$}\tabularnewline
\cline{5-8} 
 &  & $(N_{S}=4)$ & $S_{t}$ & $\uparrow$ & $\downarrow$ & $\rightarrow$ & $\leftarrow$\tabularnewline
\hline 
0 & \multirow{4}{*}{0} & 0 & $(\tilde{\mathbf{f}}_{1},\tilde{\mathbf{w}}_{1})$ & 1 & 0 & 0 & 0\tabularnewline
\cline{1-1} \cline{3-8} 
1 &  & 1 & $(\tilde{\mathbf{f}}_{1},\tilde{\mathbf{w}}_{2})$ & 0 & 0 & 1 & 0\tabularnewline
\cline{1-1} \cline{3-8} 
2 &  & 2 & $(\tilde{\mathbf{f}}_{2},\tilde{\mathbf{w}}_{2})$ & 0 & 0 & 1 & 0\tabularnewline
\cline{1-1} \cline{3-8} 
3 &  & 3 & $(\tilde{\mathbf{f}}_{3},\tilde{\mathbf{w}}_{2})$ & 1 & 0 & 0 & 0\tabularnewline
\hline 
4 & \multirow{4}{*}{1} & 0 & $(\tilde{\mathbf{f}}_{1},\tilde{\mathbf{w}}_{4})$ & 0 & 0 & 1 & 0\tabularnewline
\cline{1-1} \cline{3-8} 
5 &  & 1 & $(\tilde{\mathbf{f}}_{2},\tilde{\mathbf{w}}_{4})$ & 0 & 1 & 0 & 0\tabularnewline
\cline{1-1} \cline{3-8} 
6 &  & 2 & $(\tilde{\mathbf{f}}_{2},\tilde{\mathbf{w}}_{3})$ & 0 & 1 & 0 & 0\tabularnewline
\cline{1-1} \cline{3-8} 
7 &  & 3 & $(\tilde{\mathbf{f}}_{2},\tilde{\mathbf{w}}_{2})$ & 0 & 0 & 2 & 0\tabularnewline
\hline 
\multicolumn{1}{c}{} & \multicolumn{1}{c}{} & \multicolumn{1}{c}{$\vdots$} & \multicolumn{1}{c}{} & \multicolumn{1}{c}{} & \multicolumn{1}{c}{} & \multicolumn{1}{c}{} & \multicolumn{1}{c}{}\tabularnewline
\end{tabular}
\end{table}
The reward can be regarded as the feedback from the environment given
an action. In terms of maximizing the SINR, the reward is supposed
to be a function of SINR. Nevertheless, we only have the coupling
coefficients as measurements which suffer from noise and interference.
We therefore define the reward function as follows. According to the
received power of the coupling coefficients corresponding to the trained
beam pairs at time $t$ and $t+1$, the reward is defined, in terms
of thresholds, by functions of the received power
\begin{equation}
R_{t+1}=\begin{cases}
1, & \text{if }\frac{|y_{u,t+1}(n_{f}',n_{w}')|^{2}}{|y_{u,t}(n_{f},n_{w})|^{2}}>c_{u}\\
0, & \text{if }c_{l}<\frac{|y_{u,t+1}(n_{f}',n_{w}')|^{2}}{|y_{u,t}(n_{f},n_{w})|^{2}}\leq c_{u}\\
-1, & \text{otherwise}
\end{cases}
\end{equation}
where $(n_{f}',n_{w}')$ is the beam index pair used at time $t+1$.
Due to the noise and interference, the observations, $y_{u,t+1}(n_{f}',n_{w}')$
and $y_{u,t}(n_{f},n_{w})$, may be unreliable for determining the
reward. To reduce the uncertainty, we define a lower threshold $c_{l}$
and an upper threshold $c_{u}$. If the ratio of $|y_{u,t+1}(n_{f}',n_{w}')|^{2}$
to $|y_{u,t}(n_{f},n_{w})|^{2}$ is between $c_{l}$ and $c_{u}$,
the measurement is treated as ambiguity so that the reward is equal
to zero. A more detailed discussion about the upper and lower thresholds
is provided in Appendix \ref{sec: Conditional-Expressions-in}.

To elaborate the Q-learning-based beam selection, let us take an example
by Fig. \ref{fig: map} and Table \ref{tab:The-Q-values}. 
\begin{example}
\label{exa:When-starting-from}When starting from a state $S_{0}=(\tilde{\mathbf{f}}_{1},\tilde{\mathbf{w}}_{1})$,
one of the neighboring beam pairs $\{(\tilde{\mathbf{f}}_{1},\tilde{\mathbf{w}}_{2}),(\tilde{\mathbf{f}}_{1},\tilde{\mathbf{w}}_{6}),(\tilde{\mathbf{f}}_{2},\tilde{\mathbf{w}}_{1}),(\tilde{\mathbf{f}}_{6},\tilde{\mathbf{w}}_{1})\}$
will be explored by choosing an action from $\mathcal{A}$ according
to the state-action values, i.e., $\max_{a\in\mathcal{A}}Q(S_{0},a)$.
Since all the Q-values at $S_{0}$ are initialized to zero, an action
will be selected randomly (or according to some predefined criteria).
We assume that the action ``up'' is selected so that the next state
becomes $S_{1}=(\tilde{\mathbf{f}}_{1},\tilde{\mathbf{w}}_{2})$.
The corresponding reward and Q-value $Q(S_{0},A_{0}=\text{up})$ will
be updated accordingly, see Table \ref{tab:The-Q-values}. In the
example, we simply let the Q-values  be updated by either 0 or 1,
where a value of 1 implies that the agent chooses the action and gets
a positive reward. In Q-learning, a sequence of $N_{S}=4$ time slots
(also called steps) is defined as an \textit{episode}. Each episode
starts from a state, which could be pre-defined or determined by the
received power. Fig. \ref{fig: map} shows that the initial beam search
needs in total four episodes with starting states at $(\tilde{\mathbf{f}}_{1},\tilde{\mathbf{w}}_{1})$,
$(\tilde{\mathbf{f}}_{1},\tilde{\mathbf{w}}_{4})$, $(\tilde{\mathbf{f}}_{4},\tilde{\mathbf{w}}_{1})$,
and $(\tilde{\mathbf{f}}_{4},\tilde{\mathbf{w}}_{4})$ respectively.
In each episode, the beam probing takes $N_{S}$ time slots to update
the Q-values. When finishing the first episode, the agent starts the
next episode using beam pair $(\tilde{\mathbf{f}}_{1},\tilde{\mathbf{w}}_{4})$.
With a sufficiently large number of significant Q-values, Q-learning
will converge to the beam pair $(\tilde{\mathbf{f}}_{3},\tilde{\mathbf{w}}_{3})$
corresponding to the maximum received power.\qed

\end{example}
After the initial beam search, some beam pairs have been explored
and the beam tracking will start from the beam pair with the maximum
received power during the initial beam search, which is denoted by
$S_{M\!P}$ (i.e., the state or beam pair with respect to the maximum
power). 


According to the updated Q-values, an agent exploits what it has already
experienced in order to obtain a positive reward, but it also has
to explore the uncharted or changed environment to see if it can make
better action selections in the future. One of the challenges in reinforcement
learning is the trade-off between the exploration and exploitation.
By introducing a parameter $0<\varepsilon<1$,  an $\varepsilon$-greedy
action is obtained to better balance the exploration and exploitation:
\begin{equation}
A_{t}=\begin{cases}
\argmax_{a\in\mathcal{A}}Q(S_{t},a), & \text{with prob. }1-\varepsilon\\
\text{a random action}, & \text{with prob. }\varepsilon
\end{cases}\label{eq: e-greedy}
\end{equation}
The agent chooses the action as it believes that the action yields
the best long-term effect with probability $1-\varepsilon$. Or the
agent chooses an action uniformly at random with probability $\varepsilon$. 


The pseudocode of the Q-learning-based beam tracking algorithm is
shown in \textbf{Algorithm 1}, which includes two phases: the initial
beam search and beam tracking. The difference between these two phases
is the decision of the starting state of each episode. During the
initial beam search,  the starting state is selected from the pre-defined
sets $\mathcal{F_{I}}$ and $\mathcal{W_{I}}$. In Example \ref{exa:When-starting-from},
$\mathcal{F_{I}}=\{\tilde{\mathbf{f}}_{1},\tilde{\mathbf{f}}_{4}\}$
and $\mathcal{W_{I}}=\{\tilde{\mathbf{w}}_{1},\tilde{\mathbf{w}}_{4}\}$.
During the beam tracking, the starting state is selected according
to the maximum received power. Moreover, the selected beam pair at
time $t$ is denoted by $(\hat{\mathbf{f}}_{P,u,t},\hat{\mathbf{w}}_{P,u,t})$.
We assume that the analog beam pairs are determined at the UAV lead,
and time division duplex (TDD) technique that separates the transmit
and receive signals in the time domain can be used to inform the followers
to update their beams.

\begin{algorithm}[t]
\caption{Q-learning beam tracking for a single link.}

\begin{algorithmic}[1] 

\renewcommand{\algorithmicrequire}{\textbf{Input:}}  
\renewcommand{\algorithmicensure}{\textbf{Output:}}

\REQUIRE Observations $\{y_{u,t}(n_{f},n_{w}),t=0,\cdots,T\}$

\ENSURE Selected beam pairs $\{(\hat{\mathbf{f}}_{P,u,t},\hat{\mathbf{w}}_{P,u,t}),t=0,\cdots,T\}$

\STATE Initialize Q-table

\STATE $t=0$

\STATE \textbf{for} $i=1$ : number of episodes

\STATE\quad \textbf{if} initial beam search

\STATE\quad\quad $S_{t}\in\{(\tilde{\mathbf{f}}_{n_{f}},\tilde{\mathbf{w}}_{n_{w}})\:|\:\tilde{\mathbf{f}}_{n_{f}}\in\mathcal{F_{I}},\tilde{\mathbf{w}}_{n_{w}}\in\mathcal{W_{I}}\}$

\STATE\quad \textbf{else if} beam tracking

\STATE\quad\quad $S_{t}=S_{M\!P}$

\STATE\quad \textbf{end if}

\STATE\quad \textbf{for} $j=1$ : number $N_{S}$ of steps

\STATE\quad\quad choose $A_{t}$ and go to $S_{t+1}\equiv(\hat{\mathbf{f}}_{P,u,t+1},\hat{\mathbf{w}}_{P,u,t+1})$

\STATE\quad\quad obtain $R_{t+1}$ according to observations

\STATE\quad\quad update $Q(S_{t},A_{t})$

\STATE\quad\quad update $S_{M\!P}$ according to observations

\STATE\quad\quad $t=t+1$

\STATE\quad \textbf{end step}

\STATE \textbf{end episode}

\end{algorithmic}
\end{algorithm}

\subsection{Overhead Reduction Using Offline Q-Learning\label{subsec:Overhead-Reduction-Using}}

In Algorithm 1, the observations are available at each time slot $t$.
This implies that the beam switching and pilot transmission/reception
are executed in every time slot, which is not a well-designed manner
in the sense of system efficiency. To reduce the overhead, we reserve
all observations so that the Q-learning can execute offline. When
using past observations to obtain the rewards and update the Q-values,
we name the Q-learning algorithm \textit{offline }Q-learning. Otherwise,
it is called \textit{online }Q-learning. 

For the offline Q-learning, only the observations associated with
large received power have to be updated regularly. Therefore, at
the end of each episode, the beam pairs with respect to the maximum
received power (i.e., $S_{M\!P}$) will be chosen and employed at
the beginning of each episode in order to update the corresponding
observations. For other steps in an episode, the pilot transmission
and beam switching are not necessary unless a specific state has
not been explored.

\subsection{Beam Selection Using Q-Learning for Multiple Links\label{subsec:Beam-Selection-Using}}

The idea of Q-learning-based beam tracking for one link can be easily
extended to the case of multiple links, similar to multi-agent systems
\cite{Tan1993,Chen2019}. For multi-UAV beam probing, the lead receives
the observations from different followers simultaneously in an SDMA
manner. In this case, the members of $\mathcal{F}$ at the lead UAV's
side should not be selected repeatedly. \textcolor{red}{}As a result,
the action set in (\ref{eq: e-greedy}) has to be updated in real
time.

In each beam probing, which could be in the stage of initial beam
search or beam tracking, the Q-learning-based beam selection starts
from a follower corresponding to the maximum received power at the
moment. We further define a set $\mathcal{A}'$ that includes the
actions which will make different followers go to the same states.
Thus, the action selection given in (\ref{eq: e-greedy}) can be reformulated
as
\begin{equation}
A_{t}=\begin{cases}
\argmax_{a\in\mathcal{A\backslash A}'}Q(S_{t},a), & \text{with prob. }1-\varepsilon\\
\text{randomly selected from }\mathcal{A\backslash A}', & \text{with prob. }\varepsilon
\end{cases}\label{eq: e-greedy-1}
\end{equation}
After making the decision about the next state for a follower, the
lead has to update $\mathcal{A}'$ accordingly.

\section{Digital Beamforming}

In the previous section, we use Q-learning to find the members of
sets $\mathcal{I_{F}}_{\!,t}$ and $\mathcal{I_{W}}_{\!,t}$\textcolor{red}{
}in the problem (\ref{eq: max}). However, the selected beam pairs
may not be the optimal solution to the problem for the reasons that
(i) Q-learning usually only provides a good enough solution\footnote{Q-learning uses experience to solve a prediction problem, which can
be viewed as a Monte Carlo method. } and (ii) the digital beamformer weights are not taken into account
during the procedure of analog beam selection. In the sense of hybrid
beamforming, a better solution should be the one whose linear combination
with the digital weights leading to the maximum SINR. This issue can
be solved by keeping more than one promising members with large received
power in $\mathcal{I_{F}}_{\!,t}$ and $\mathcal{I_{W}}_{\!,t}$ \cite{Chiang2018_JSTSP}.
We use Example \ref{exa:If-two-selected} to explain the idea.
\begin{example}
\label{exa:If-two-selected}Two selected beam pairs with large received
power for each follower are collected in the following two sets:
\[
\{[\tilde{\mathbf{f}}_{P,1},\tilde{\mathbf{f}}_{P,2},\tilde{\mathbf{f}}_{P,3}],[\tilde{\mathbf{f}}_{P,1},\tilde{\mathbf{f}}_{P,3},\tilde{\mathbf{f}}_{P,4}]\}
\]
and
\[
\{[\tilde{\mathbf{w}}_{P,1},\tilde{\mathbf{w}}_{P,1},\tilde{\mathbf{w}}_{P,2}],[\tilde{\mathbf{w}}_{P,2},\tilde{\mathbf{w}}_{P,3},\tilde{\mathbf{w}}_{P,4}]\}.
\]
Given these two sets, we can generate all the members of $\mathcal{I_{F}}_{\!,t}$
and $\mathcal{I_{W}}_{\!,t}$, given by
\[
\mathcal{I_{F}}_{\!,t}=\{\underset{\footnotesize\begin{array}{c}
\text{the }1^{\text{st}}\text{ candiate}\\
\text{for }\mathbf{F}_{P,t}
\end{array}}{\underbrace{[\tilde{\mathbf{f}}_{P,1},\tilde{\mathbf{f}}_{P,2},\tilde{\mathbf{f}}_{P,3}]}},[\tilde{\mathbf{f}}_{P,1},\tilde{\mathbf{f}}_{P,2},\tilde{\mathbf{f}}_{P,4}],[\tilde{\mathbf{f}}_{P,1},\tilde{\mathbf{f}}_{P,3},\tilde{\mathbf{f}}_{P,4}]\}
\]
which has a cardinality of 3 because the members of $\mathcal{F}$
at lead UAV should not be selected repeatedly, and the other set can
be represented by
\begin{multline*}
\mathcal{I_{W}}_{\!,t}=\{\underset{\footnotesize\begin{array}{c}
\text{the }1^{\text{st}}\text{ candiate}\\
\text{for }\mathbf{W}_{P,t}
\end{array}}{\underbrace{[\tilde{\mathbf{w}}_{P,1},\tilde{\mathbf{w}}_{P,1},\tilde{\mathbf{w}}_{P,2}]}},[\tilde{\mathbf{w}}_{P,1},\tilde{\mathbf{w}}_{P,1},\tilde{\mathbf{w}}_{P,4}],\cdots,\\{}
[\tilde{\mathbf{w}}_{P,2},\tilde{\mathbf{w}}_{P,3},\tilde{\mathbf{w}}_{P,4}]\}
\end{multline*}
which has a cardinality of 8. In this example, given the above $\mathcal{I_{F}}_{\!,t}$
and $\mathcal{I_{W}}_{\!,t}$, we have to evaluate a total of 24 combinations
with their digital weights to maximize the SINR.\qed
\end{example}

The above-mentioned idea is different from the work represented in
\cite{Chen2018} that keeps candidates in subspace. In our opinion,
the better solution is supposed to keep candidates with large received
power because the idea in \cite{Chen2018} only takes into account
the main lobes of analog beams, while the proposed method considers
both the main and side lobes. 

\subsection{Digital Weight Optimization}

To simplify the following descriptions of digital beamforming, we
assume that $\mathcal{I_{F}}_{\!,t}$ and $\mathcal{I_{W}}_{\!,t}$
only include one member respectively, i.e.,
\begin{align}
\mathcal{I_{F}}_{\!,t} & =\{\hat{\mathbf{F}}_{P,t}=[\hat{\mathbf{f}}_{P,1,t},\cdots,\hat{\mathbf{f}}_{P,U,t}]\}\nonumber \\
\mathcal{I_{W}}_{\!,t} & =\{\hat{\mathbf{W}}_{P,t}=[\hat{\mathbf{w}}_{P,1,t},\cdots,\hat{\mathbf{w}}_{P,U,t}]\}.
\end{align}
In the numerical results, we will provide more discussion about the
idea. Given $\hat{\mathbf{F}}_{P,t}$ and $\hat{\mathbf{W}}_{P,t}$,
the hybrid beamforming problem (\ref{eq: max}) becomes a digital
beamforming problem subject to the power constraint, which can be
formulate as
\begin{equation}
\begin{gathered}\sum_{u=1}^{U}\,\underset{\mathbf{f}_{B,u,t}}{\max}\,\frac{\hat{P}_{S,u,t}}{\hat{P}_{I,u,t}+\sigma_{n}^{2}}\\
\text{s.t. }\mathbf{f}_{B,u,t}^{H}\hat{\mathbf{F}}_{P,t}^{H}\hat{\mathbf{F}}_{P,t}\mathbf{f}_{B,u,t}=1\;\forall u
\end{gathered}
\label{eq: max-1}
\end{equation}
where $t=1,\cdots,T$. The signal and interference power are subject
to the selected analog beams
\begin{align}
\hat{P}_{S,u,t} & \triangleq P_{S,u,t}\,\vline\,_{\mathbf{F}_{P,t}=\hat{\mathbf{F}}_{P,t},\mathbf{W}_{P,t}=\hat{\mathbf{W}}_{P,t}},\\
\hat{P}_{I,u,t} & \triangleq P_{I,u,t}\,\vline\,_{\mathbf{F}_{P,t}=\hat{\mathbf{F}}_{P,t},\mathbf{W}_{P,t}=\hat{\mathbf{W}}_{P,t}}.
\end{align}

To satisfy the power constraint on the combiner, one can define $U$
unit vectors $\{\mathbf{x}_{u}\,|\,\left\Vert \mathbf{x}_{u}\right\Vert _{2}=1,u=1,\cdots,U\}$
that obey the relation \cite{Chiang2018_JSTSP}
\begin{equation}
\mathbf{f}_{B,u,t}=(\hat{\mathbf{F}}_{P,t}^{H}\hat{\mathbf{F}}_{P,t})^{-0.5}\mathbf{x}_{u}.
\end{equation}
Upon replacing $\mathbf{f}_{B,u,t}$ with $(\hat{\mathbf{F}}_{P,t}^{H}\hat{\mathbf{F}}_{P,t})^{-0.5}\mathbf{x}_{u}$
in the problem, the received signal and interference power can be
written by
\begin{align}
\hat{P}_{S,u,t} & =\left|\mathbf{x}_{u}^{H}(\hat{\mathbf{F}}_{P,t}^{H}\hat{\mathbf{F}}_{P,t})^{-0.5}\hat{\mathbf{F}}_{P,t}^{H}\mathbf{H}_{u,t}\hat{\mathbf{w}}_{P,u,t}\right|^{2},\label{eq: P_S}\\
\hat{P}_{I,u,t} & =\sum_{i=1,i\neq u}^{U}\left|\mathbf{x}_{u}^{H}(\hat{\mathbf{F}}_{P,t}^{H}\hat{\mathbf{F}}_{P,t})^{-0.5}\hat{\mathbf{F}}_{P,t}^{H}\mathbf{H}_{i,t}\hat{\mathbf{w}}_{P,i,t}\right|^{2}.\label{eq: P_I}
\end{align}
Then, we can find that the problem (\ref{eq: max-1}) is equivalent
to seeking vectors $\mathbf{x}_{1},\cdots,\mathbf{x}_{U}$ that maximize
the SINR for $U$ followers. As a result, the maximization problem
(\ref{eq: max-1}) can be reformulated as
\begin{equation}
\sum_{u=1}^{U}\:\underset{\mathbf{x}_{u}}{\max}\:\frac{\hat{P}_{S,u,t}}{\hat{P}_{I,u,t}+\sigma_{n}^{2}}.
\end{equation}

\subsection{SINR Approximation Using Coupling Coefficients}

In (\ref{eq: P_S}) and (\ref{eq: P_I}), the couplings of the channel
and analog beams, such as $\hat{\mathbf{F}}_{P,t}^{H}\mathbf{H}_{u,t}\hat{\mathbf{w}}_{P,u,t}$
and $\hat{\mathbf{F}}_{P,t}^{H}\mathbf{H}_{i,t}\hat{\mathbf{w}}_{P,i,t}$,
can be viewed as effective channel vectors. Since the observations,
given in (\ref{eq: y}), are the coupling of the channel and one analog
beam pair, we can use them to construct the estimates of effective
channel vectors, defined by
\begin{align}
\hat{\mathbf{h}}_{E,u,t} & =\hat{\mathbf{F}}_{P,t}^{H}\mathbf{H}_{u,t}\hat{\mathbf{w}}_{P,u,t}+z_{t}\nonumber \\
 & =\underset{\footnotesize\begin{array}{c}
\text{The entries of }\hat{\mathbf{h}}_{E,u,t}\text{ can be}\\
\text{obtained from }\{y_{u,t}(n_{f},n_{w})\:\forall u\}
\end{array}}{\underbrace{\left[\begin{array}{c}
\hat{\mathbf{f}}_{P,1,t}^{H}\mathbf{H}_{u,t}\hat{\mathbf{w}}_{P,u,t}+z_{t}\\
\vdots\\
\hat{\mathbf{f}}_{P,U,t}^{H}\mathbf{H}_{u,t}\hat{\mathbf{w}}_{P,u,t}+z_{t}
\end{array}\right]}},
\end{align}
and
\begin{equation}
\hat{\mathbf{h}}_{E,i,t}=\hat{\mathbf{F}}_{P,t}^{H}\mathbf{H}_{i,t}\hat{\mathbf{w}}_{P,i,t}+z_{t},
\end{equation}
where the entries of $\hat{\mathbf{h}}_{E,i,t}$ can be obtained from
$\{y_{u,t}(n_{f},n_{w})\:\forall u\}$ as well. The collected observations
suffice to generate the estimates of $\hat{P}_{S,u,t}$ and $\hat{P}_{I,u,t}+\sigma_{n}^{2}$
represented by

\begin{align}
\hat{P}_{S,u,t} & \approx\left|\mathbf{x}_{u}^{H}(\hat{\mathbf{F}}_{P,t}^{H}\hat{\mathbf{F}}_{P,t})^{-0.5}\hat{\mathbf{h}}_{E,u,t}\right|^{2}\nonumber \\
 & =\mathbf{x}_{u}^{H}\underset{\triangleq\mathbf{A}_{u,t}}{\underbrace{(\hat{\mathbf{F}}_{P,t}^{H}\hat{\mathbf{F}}_{P,t})^{-0.5}\hat{\mathbf{h}}_{E,u,t}\hat{\mathbf{h}}_{E,u,t}^{H}(\hat{\mathbf{F}}_{P,t}^{H}\hat{\mathbf{F}}_{P,t})^{-0.5}}}\mathbf{x}_{u}\nonumber \\
 & =\mathbf{x}_{u}^{H}\mathbf{A}_{u,t}\mathbf{x}_{u}\label{eq: Ps_hat}
\end{align}
and{\scriptsize{}
\begin{flalign}
 & \hat{P}_{I,u,t}+\sigma_{n}^{2}\nonumber \\
 & \approx\sum_{i=1,i\neq u}^{U}\left|\mathbf{x}_{u}^{H}(\hat{\mathbf{F}}_{P,t}^{H}\hat{\mathbf{F}}_{P,t})^{-0.5}\hat{\mathbf{h}}_{E,i,t}\right|^{2}+\sigma_{n}^{2}\nonumber \\
 & =\mathbf{x}_{u}^{H}\underset{\triangleq\mathbf{B}_{u,t}}{\underbrace{\left(\sum_{i=1,i\neq u}^{U}(\hat{\mathbf{F}}_{P,t}^{H}\hat{\mathbf{F}}_{P,t})^{-0.5}\hat{\mathbf{h}}_{E,i,t}\hat{\mathbf{h}}_{E,i,t}^{H}(\hat{\mathbf{F}}_{P,t}^{H}\hat{\mathbf{F}}_{P,t})^{-0.5}+\sigma_{n}^{2}\mathbf{I}_{U}\right)}}\mathbf{x}_{u}\nonumber \\
 & =\mathbf{x}_{u}^{H}\mathbf{B}_{u,t}\mathbf{x}_{u}\label{eq: Pi_hat}
\end{flalign}
}{\scriptsize \par}

Using (\ref{eq: Ps_hat}) and (\ref{eq: Pi_hat}), the SINR for follower
$u$ conditional on $\mathbf{W}_{P,t}=\hat{\mathbf{W}}_{P,t}$ and
$\mathbf{F}_{P,t}=\hat{\mathbf{F}}_{P,t}$ can be approximated by
the following equation
\begin{align}
\frac{\hat{P}_{S,u,t}}{\hat{P}_{I,u,t}+\sigma_{n}^{2}} & \approx\frac{\mathbf{x}_{u}^{H}\mathbf{A}_{u,t}\mathbf{x}_{u}}{\mathbf{x}_{u}^{H}\mathbf{B}_{u,t}\mathbf{x}_{u}}.\label{eq: est_SINR}
\end{align}
Using the property that $\mathbf{B}_{u,t}$ is a positive definite
matrix, the optimal solution of $\mathbf{x}_{u}$ that attains the
maximum SINR can be stated as follows (also see Appendix \ref{sec:Derivation-of-()}):
\begin{align}
\mathbf{x}_{u}^{\star} & =\underset{\mathbf{x}_{u}}{\arg\,\max}\:\frac{\mathbf{x}_{u}^{H}\mathbf{A}_{u,t}\mathbf{x}_{u}}{\mathbf{x}_{u}^{H}\mathbf{B}_{u,t}\mathbf{x}_{u}}\nonumber \\
 & =\frac{\mathbf{B}_{u,t}^{-0.5}\mathbf{e}_{\max}(\mathbf{B}_{u,t}^{-0.5}\mathbf{A}_{u,t}\mathbf{B}_{u,t}^{-0.5})}{\left\Vert \mathbf{B}_{u,t}^{-0.5}\mathbf{e}_{\max}(\mathbf{B}_{u,t}^{-0.5}\mathbf{A}_{u,t}\mathbf{B}_{u,t}^{-0.5})\right\Vert _{2}},\label{eq: x_hat}
\end{align}
where $\mathbf{e}_{\max}(\mathbf{B}_{u,t}^{-0.5}\mathbf{A}_{u,t}\mathbf{B}_{u,t}^{-0.5})$
is the eigenvector of $\mathbf{B}_{u,t}^{-0.5}\mathbf{A}_{u,t}\mathbf{B}_{u,t}^{-0.5}$
corresponding to the maximum eigenvalue. In the same manner, given
$\mathbf{A}_{u,t}$ and $\mathbf{B}_{u,t}$ for all $u$, we have
the optimal solution of $\mathbf{x}_{u}$, $u=1,\cdots,U$. The corresponding
estimated digital beamformer weights are therefore given by
\begin{align}
\hat{\mathbf{F}}_{B,t} & =[\hat{\mathbf{f}}_{B,1,t},\cdots,\hat{\mathbf{f}}_{B,U,t}]\nonumber \\
 & =(\hat{\mathbf{F}}_{P,t}^{H}\hat{\mathbf{F}}_{P,t})^{-0.5}[\mathbf{x}_{1}^{\star},\cdots,\mathbf{x}_{U}^{\star}].\label{eq: DBF}
\end{align}

The digital weights represented in (\ref{eq: DBF}) are derived from
the constraint that the variance of the combined noise signal is still
AWGN. When concerning $\mathbf{F}_{B,t}$ acting as part of the precoder
for data transmission (i.e., sending signals from the lead to followers),
the power constraint could be $||\mathbf{F}_{P,t}\mathbf{F}_{B,t}||_{F}=U$
\cite{Alkhateeb2015a}, which leads to zero-forcing (ZF) digital beamforming.

\section{Simulation Results}

\begin{figure}[t]
\includegraphics[scale=0.6]{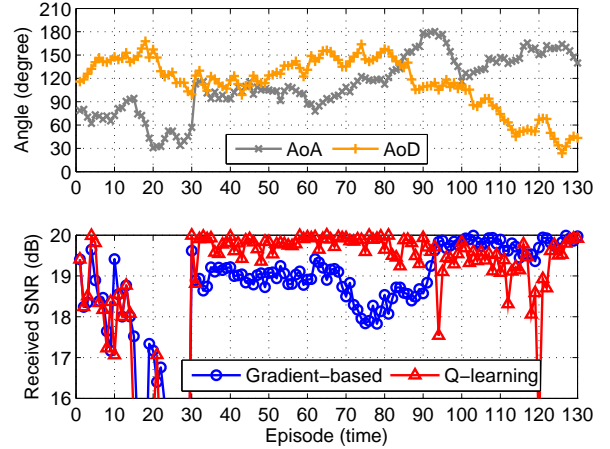}

\caption{An typical example of the received SNR using the Q-learning-based
and gradient-based beam tracking methods for one link ($U=1$) with
$\sigma_{\lambda}^{2}=16$. In this example, we assume fixed elevation
angles $\theta_{A,u,t}=\theta_{D,u,t}=15^{\circ}\:\forall t$. Compared
with the gradient-based method, the Q-learning-based beam tracking
is robust to the large variance of angle.\label{fig:one-relization}}
\end{figure}
\begin{figure*}[t]
\begin{centering}
\includegraphics{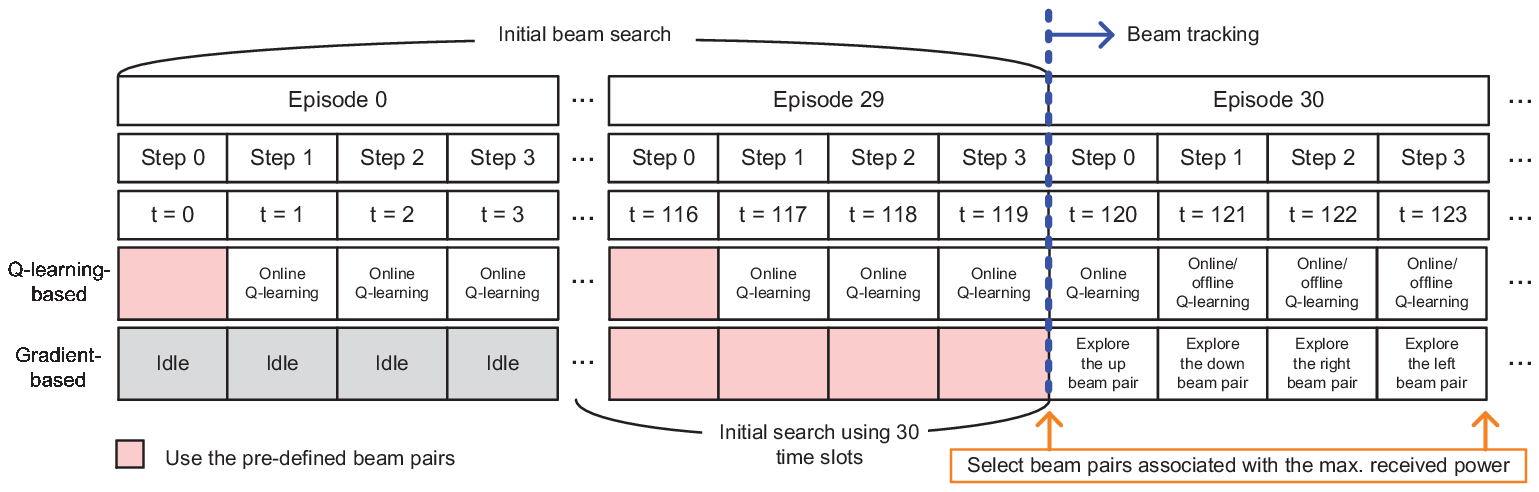}
\par\end{centering}
\caption{Time frame defined by the episode and step. In the Q-learning-based
beam tracking, we can use offline Q-learning in some steps for the
purpose of overhead reduction (introduced in Subsections \ref{subsec:Resilience-and-Efficiency}
and \ref{subsec:SINR-Maximization-Using}).\label{fig:Beam-tracking-time}}
\end{figure*}
In this section, we numerically illustrate the multi-UAV beamforming
performance in highly dynamic environments, while each result at
a time slot averages 1000 trials. The system parameters in the simulations
are listed as follows.
\begin{itemize}
\item The lead connects to $U=3$ followers using SDMA at the same time
and same frequency. The number of antennas $N=16$ $(4\times4)$,
and the $\text{SINR}=20\text{ dB}$ in the simulations for each follower
is given by $\frac{|\rho_{u}|^{2}}{\sigma_{z}^{2}}$, where $|\rho_{u}|^{2}$
is the average receive power for follower $u$ and $\Sigma_{u=1}^{U}|\rho_{u}|^{2}=1$. 
\item In the codebooks, the candidates for azimuth angles $\phi_{n_{f}}$
and $\phi_{n_{w}}$ are $\{15^{\circ}+n\cdot30^{\circ}\}_{n=0}^{11}$,
and the candidates for elevation angles $\theta_{n_{f}}$ and $\theta_{n_{w}}$
are $\{15^{\circ}+n\cdot30^{\circ}\}_{n=0}^{2}$. 
\item The Q-learning parameters include the learning rate $\alpha=0.5$,
discount factor $\gamma=0.5$, probability of $\varepsilon$-greedy
action $\varepsilon=0.1$, upper threshold $c_{u}=1.1$, and lower
threshold $c_{l}=0.9$. 
\item The random walk channel model has normally distributed disturbance
$\lambda_{i}\sim\mathcal{N}(0,\sigma_{\lambda}^{2})$, where $\sigma_{\lambda}^{2}=4,16$. 
\end{itemize}
According to the number of all potential steering angles, the size
of codebook $\mathcal{F}$ at the lead should be 36. To alleviate
the loading at the lead and speed up the convergence and learning
rate, we group the followers into three zones in elevation angle (i.e.,
$0^{\circ}-30^{\circ}$, $30^{\circ}-60^{\circ}$, and $60^{\circ}-90^{\circ}$),
and each zone has three followers. Due to the space limitation, we
only show the simulation results with three followers in the zone
of elevation angle between $0^{\circ}$ and $30^{\circ}$, and the
codebook size of $\mathcal{F}$ becomes $N_{F}=12$, where the $12$
candidates all have the same elevation angle $\theta_{n_{f}}=15^{\circ}$.

\subsection{Q-Learning and Gradient-Based Beam Tracking Methods\label{subsec:Q-Learning-and-Gradient-Based}}

The first numerical result of the beam tracking in Fig. \ref{fig:one-relization}
is described by an example of the performance comparison of the proposed
Q-learning and reference gradient-based tracking methods \cite{Kadur2018}.
We use one realization of the time-varying AoA and AoD with $\sigma_{\lambda}^{2}=16$
to explain the difference between these two methods. 

\begin{figure}[t]
\begin{centering}
\includegraphics[scale=0.6]{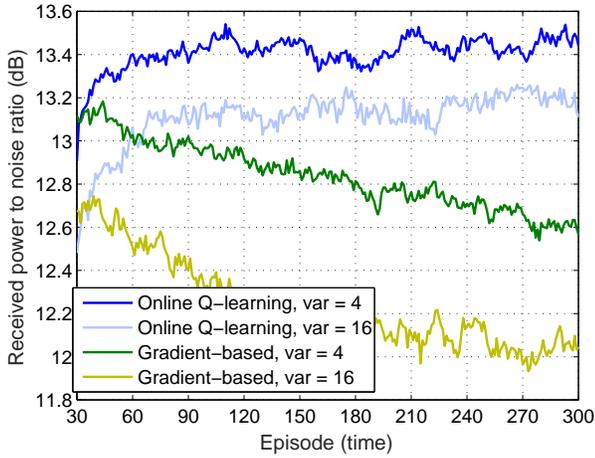}
\par\end{centering}
\caption{Sum of the received power from $U=3$ followers to noise ratio using
the proposed Q-learning and reference gradient-based beam tracking
methods with the variance of AoAs and AoDs $\sigma_{\lambda}^{2}=4,16$.
Compared with the gradient-based method, the Q-learning-based beam
tracking can provide stable link quality over time.\label{fig:comparison} }
\end{figure}
At the beginning, both methods implement the initial beam search in
the first 30 episodes or time slots\footnote{The 30 beam pairs for the initial beam search are uniformly chosen
from a total of $N_{F}\times N_{W}=12\times36=432$ potential beam
pairs.}, where the time frame is sketched in Fig. \ref{fig:Beam-tracking-time}.
In the time frame, we assume that each episode includes 4 steps so
that the gradient-based approach can evaluate the 4 neighboring beam
pairs in an episode during the beam tracking. The gradient-based method
uses 30 \textit{time slots} to implement the initial beam search,
while Q-learning method uses 30\textit{ episodes} to implement the
initial beam search and update the Q-values at each step. After the
initial beam search, the beam tracking starts from Episode 30 with
the state corresponding to the maximum received power obtained during
the previous 30 episodes. The Q-learning method during the beam tracking
may adopt online or offline Q-learning. To fairly compare with the
reference method which gets the latest observations at each time slot,
we use \textit{online} Q-learning for all the steps in each episode
to evaluate the proposed method in Fig. \ref{fig:one-relization}.

In Fig. \ref{fig:one-relization}, from Episode 30 to 80, the Q-learning-based
beam tracking explored the range of AoA within $[60^{\circ},120^{\circ}]$
and the range of AoD within $[90^{\circ},170^{\circ}]$ using the
beams steering to $\phi_{n_{f}}=75^{\circ}$, $105^{\circ}$ and $\phi_{n_{w}}=105^{\circ},135^{\circ},165^{\circ}$,
respectively.\footnote{The beamwidth is around $30^{\circ}$; ideally the beam switching
occurs when AoA/AoD changes at $30^{\circ},60^{\circ},\cdots,180^{\circ}$
in Fig. \ref{fig:one-relization}.} Q-learning records all the experience acquired during this time
in the Q-table so that the agent uses current observations and the
experience to predict the next beam pair. In such a way, it can stably
track the appropriate beams. After Episode 80, there are probably
not many data corresponding to the beam pairs with $\phi_{n_{f}}=135^{\circ},165^{\circ}$
or $\phi_{n_{w}}=45^{\circ},75^{\circ}$; therefore, it needs some
time to update the Q-values as reference in the future. Next, let
us look at the performance of the reference gradient-based scheme.
The delimma of gradient-based scheme is that it may get trapped into
a local optimum and could only get out from it when AoA or AoD changes
significantly. Q-learning method also finds the local optimal solution
sometimes, but appropriate $\varepsilon$-greedy random actions can
solve this problem. Moreover, compared with the gradient-based method,
Q-learning has a \textit{global} map (i.e., the Q-table), which provides
useful information for beam tracking.

The performance comparison of the proposed and reference methods that
support $U=3$ followers simultaneously is shown in Fig. \ref{fig:comparison},
where the received power is captured at the end of each episode as
described in Fig. \ref{fig:Beam-tracking-time}. Compared with the
gradient-based method, the Q-learning-based scheme works stable over
time, even when the variance $\sigma_{\lambda}^{2}$ of AoAs and AoDs
is large. In terms of high-resilience demand for the multi-UAV system,
the numerical results of the proposed method show that balancing the
exploration and exploitation can outperform the one using exploration
only. Although Q-learning needs some space and efforts to record the
experience in the Q-table,  it makes actions depending on not only
current observations but also the experience and rewards so that the
performance is not completely dominated by the current observations,
while the gradient-based method totally relies on them.

\subsection{Resilience and Efficiency of Offline Q-learning-Based Beam Tracking\label{subsec:Resilience-and-Efficiency}}

\begin{figure}[t]
\begin{centering}
\includegraphics[scale=0.6]{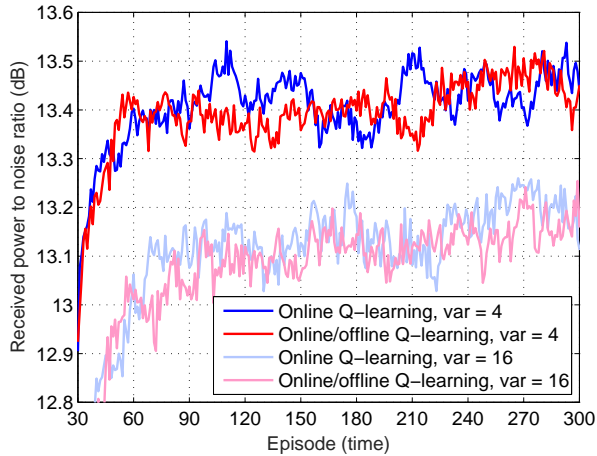}
\par\end{centering}
\caption{Sum of the received power from $U=3$ followers to noise ratio using
the online and online/offline Q-learning-based beam tracking methods
with $\sigma_{\lambda}^{2}=4,16$. The curves of online Q-learning
in this figure and Fig. \ref{fig:comparison} are identical. According
to the results of overhead reduction in Fig. \ref{fig:Reduced-overhead-of},
using one or at most two steps per episode to track the beams is enough
to maintain the link quality.\label{fig:Sum-of-the} }
\end{figure}

\begin{figure}[t]
\begin{centering}
\includegraphics[scale=0.6]{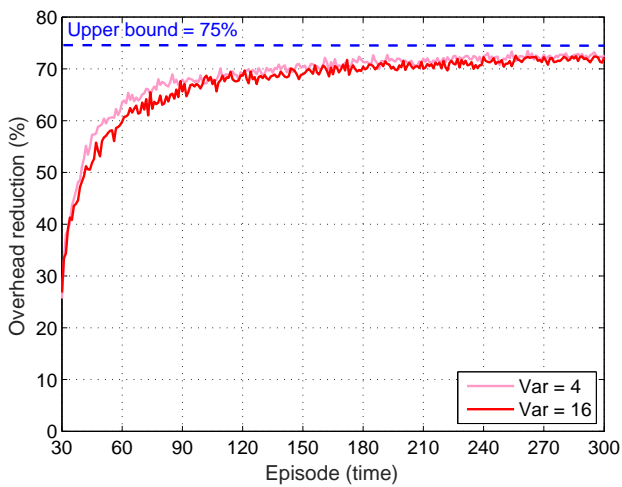}
\par\end{centering}
\caption{Reduced overhead of pilot transmission using the online/offline Q-learning-based
beam tracking with $\sigma_{\lambda}^{2}=4,16$. In an episode, the
first step is always used for pilot transmission so that the upper
bound is 75\%, while the other three steps may or may not be used
for pilot transmission, depending on whether the corresponding states
are explored or not.\label{fig:Reduced-overhead-of}}
\end{figure}

In the previous subsection, we use \textit{online} Q-learning to implement
the beam tracking at all the steps in each episode in order to compare
with the reference method. From the results shown in Fig. \ref{fig:comparison},
we observe that online Q-learning provides stable link quality that
can meet the high-resilience requirement for highly dynamic multi-UAV
environments, but it also means that all the resources are used as
pilot signals. From the perspective of system efficiency, it is inefficient
design. Essentially, the trade-off between system efficiency and resilience
has to be considered together. As a result, we introduce \textit{offline}
Q-learning that uses past observations to implement the beam tracking.

At the first step in each episode during the beam tracking, we let
the followers transmit the pilot signals using the selected beam pairs,
determined in the previous episodes, to update the observations. Therefore,
the first step always adopts online Q-learning. In the rest three
steps, the next state $S_{t+1}$ (decided by Q-learning) may or may
not be explored in the previous episodes. If the state was not explored,
the agent still adopts online Q-learning in order to get the corresponding
reward as well as Q-value. Instead, if the state was explored, Q-learning
can use past observations to implement the beam tracking, which is
offline Q-learning. However, we are not sure whether the next states
in the rest three steps were explored. Therefore, the agent may adopt
online or offline Q-learning, i.e., the case \textit{online/offline}
Q-learning in Figs. \ref{fig:Beam-tracking-time} and \ref{fig:Sum-of-the}.
In such a design, the upper bound of the reduced overhead of pilot
transmission is 75\%, see Fig. \ref{fig:Reduced-overhead-of}, since
one of the four steps in an episode is dedicated to pilot transmission.

Figs. \ref{fig:Sum-of-the} and \ref{fig:Reduced-overhead-of} show
the comparison of the online and online/offline Q-learning-based beam
tracking methods. The simulations results provide some interesting
insights. After a certain time of exploration, the overhead of pilot
transmission could be reduced up to 72\% without any loss of performance.
This implies that the experience stored in the Q-table provides enough
information to solve the prediction problem, which is an advantage
of machine learning.

\subsection{SINR Maximization Using Digital Weights\label{subsec:SINR-Maximization-Using}}

\begin{figure}[t]
\includegraphics[scale=0.6]{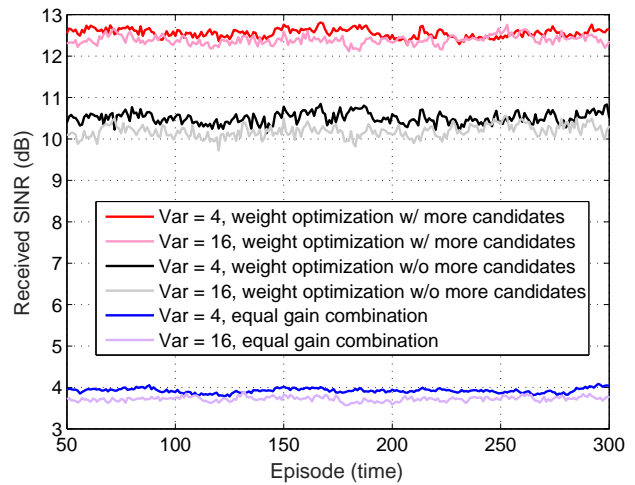}\caption{Achievable SINR that uses equal gain or optimal weights to combine
$U=3$ analog beam pairs with $\sigma_{\lambda}^{2}=4,16$. With and
without the optimal digital weights, the difference in received SINR
is 6.5 dB. With more than one candidate for the analog beam pairs,
the received SINR can be further improved by at least 2 dB. \label{fig:Achievable-SINR-of}}
\end{figure}

We use SDMA to support multi-UAV communications at the price of spatial-domain
interference. The goal of the digital beamforming is to minimize the
interference plus noise given the selected analog beams. The performance
is shown in Fig. \ref{fig:Achievable-SINR-of}, where the received
SINR is obtained using $\hat{\mathbf{F}}_{P,t}$, $\hat{\mathbf{W}}_{P,t}$,
and $\hat{\mathbf{F}}_{B,t}$.

First, the curves \textit{equal gain combination} do not take into
account more than one candidate for the analog beam pairs (i.e., both
$\mathcal{I_{F}}_{\!,t}$ and $\mathcal{I_{W}}_{\!,t}$ have only
one member respectively), and we let the digital beamforming equal
to the $U\times U$ identity matrix (i.e., let $\hat{\mathbf{F}}_{B,t}=\mathbf{I}_{U}$).
Without trying to minimize the interference, the achievable SINR is
only around 4 dB.  When we design the digital weights to minimize
the interference plus noise, see curves \textit{w/o more candidates},
the SINR gain can be increased by around 6.5 dB. If we keep more candidates
for the analog beams after the beam selection at the end of each episode
(see Example \ref{exa:If-two-selected}), the SINR can be further
improved. In curves \textit{w/ more candidates}, reserving two candidates
for each beam pair after the beam selection can have 2 dB gain in
SINR. The reason is that the selected analog beams based on the received
power do not ensure that the hybrid beamforming achieves the maximization
SINR, even with the corresponding optimal digital weights.

Comparing the curves with low and high changes of the angle variables
corresponding to low and high speeds of the UAVs in Fig. \ref{fig:Achievable-SINR-of},
we can see that the maximum difference in SINR is less than 0.5 dB.
It shows that the proposed Q-learning-based hybrid beamforming is
quite robust to the large variance of AoAs and AoDs.

\section{Conclusion}

This paper solved a highly dynamic multi-UAV hybrid beamforming problem
that is usually characterized by a high-resilience constraint. To
meet the constraint, we apply Q-learning method to mmWave hybrid beamforming
systems. Moreover, in a dynamic environment, how to efficiently obtain
and use the observations matters to the beamforming performance. In
the proposed analog beam tracking approach, we use current and past
observations together with the designed rewards to solve the prediction
problem. The numerical results show that the proposed method significantly
increases the efficiency of data transmission and beam switching.
To optimally combine the analog beams in a manner of SINR maximization,
we present the solution of digital weights using the coupling coefficients
given the selected beams. The solution can be simply extended to the
case with more candidates for analog beams to further improve the
received SINR.

\appendices{}

\section{Design of Upper and Lower Thresholds ($c_{u},c_{l}$)\label{sec: Conditional-Expressions-in}}

In the observation equation (\ref{eq: y}), given a channel matrix,
the estimate of the power of signal $y_{u,t}(n_{f},n_{w})$ can be
represented by
\begin{flalign}
\left|y_{u,t}(n_{f},n_{w})\right|^{2} & =\left|\tilde{\mathbf{f}}_{n_{f}}^{H}\mathbf{H}_{u,t}\tilde{\mathbf{w}}_{n_{w}}+z_{t}\right|^{2}\nonumber \\
 & =\left|\tilde{\mathbf{f}}_{n_{f}}^{H}\mathbf{H}_{u,t}\tilde{\mathbf{w}}_{n_{w}}\right|^{2}+\varepsilon_{u}(n_{w},n_{f})+\zeta,
\end{flalign}
{\small{}w}here $|\tilde{\mathbf{f}}_{n_{f}}^{H}\mathbf{H}_{u,t}\tilde{\mathbf{w}}_{n_{w}}|^{2}$
is a constant based on the given channel state, and the other two
terms are given as follows. First, $\varepsilon_{u}(n_{w},n_{f})$
follows a normal distribution with mean zero and variance $2\sigma_{z}^{2}|\tilde{\mathbf{f}}_{n_{f}}^{H}\mathbf{H}_{u,t}\tilde{\mathbf{w}}_{n_{w}}|^{2}$
given by
\begin{align}
 & \varepsilon_{u}(n_{w},n_{f})\nonumber \\
 & =2\,\mathfrak{R}\left(\tilde{\mathbf{f}}_{n_{f}}^{H}\mathbf{H}_{u,t}\tilde{\mathbf{w}}_{n_{w}}\right)\mathfrak{R}\left(z_{t}\right)+2\,\mathfrak{I}\left(\tilde{\mathbf{f}}_{n_{f}}^{H}\mathbf{H}_{u,t}\tilde{\mathbf{w}}_{n_{w}}\right)\mathfrak{I}\left(z_{t}\right)\nonumber \\
 & \sim\mathcal{N}\left(0,2\sigma_{z}^{2}\left|\tilde{\mathbf{f}}_{n_{f}}^{H}\mathbf{H}_{u,t}\tilde{\mathbf{w}}_{n_{w}}\right|^{2}\right),
\end{align}
and $\zeta$ follows a gamma distribution with shape parameter $1$
and scale parameter $\sigma_{z}^{2}$:
\begin{align}
\zeta & =\mathfrak{R}\left(z_{t}\right)^{2}+\mathfrak{I}\left(z_{t}\right)^{2}\sim\Gamma\left(1,\sigma_{z}^{2}\right).
\end{align}

\begin{figure}[t]
\begin{centering}
\includegraphics[scale=0.55]{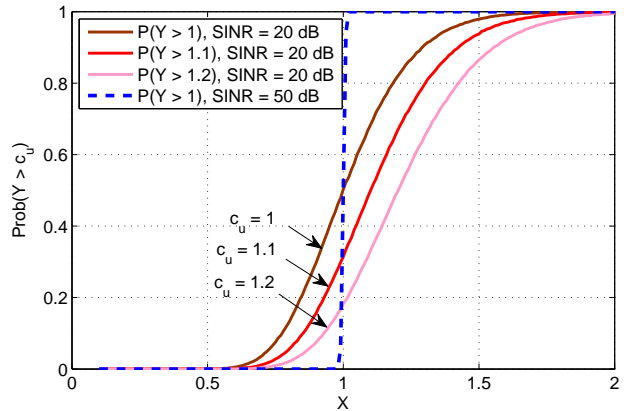}
\par\end{centering}
\caption{The probabilities that the estimate of $X$ is greater than the upper
threshold $c_{u}$. Increasing the value of the upper threshold $c_{u}$
can reduce the probability that the agent get a fail reward at $\text{SINR}=20\text{ dB}$.\label{fig:The-probability-of}}

\end{figure}

Due to the fact that it is not able to obtain a closed-form expression
for the density function of $\varepsilon_{u}(n_{w},n_{f})+\zeta$,
we use a Monte Carlo method to find appropriate upper and lower thresholds.
First, let us define the ratio of the power of coupling coefficients
at time $t+1$ and $t$ by
\begin{equation}
X=\frac{|\tilde{\mathbf{f}}_{n_{f}'}^{H}\mathbf{H}_{u,t+1}\tilde{\mathbf{w}}_{n_{w}'}|^{2}}{|\tilde{\mathbf{f}}_{n_{f}}^{H}\mathbf{H}_{u,t}\tilde{\mathbf{w}}_{n_{w}}|^{2}},
\end{equation}
where $(n_{f}',n_{w}')$ is the beam index pair used at time $t+1$.
In addition, the ratio of the received power at time $t+1$ and $t$
is given by
\begin{align}
Y & =\frac{|y_{u,t+1}(n_{f}',n_{w}')|^{2}}{|y_{u,t}(n_{f},n_{w})|^{2}}\nonumber \\
 & =\frac{|\tilde{\mathbf{f}}_{n_{f}'}^{H}\mathbf{H}_{u,t+1}\tilde{\mathbf{w}}_{n_{w}'}|^{2}+\varepsilon_{u}(n_{w}',n_{f}')+\zeta_{1}}{|\tilde{\mathbf{f}}_{n_{f}}^{H}\mathbf{H}_{u,t}\tilde{\mathbf{w}}_{n_{w}}|^{2}+\varepsilon_{u}(n_{w},n_{f})+\zeta_{2}},
\end{align}
where $\zeta_{1}$ and $\zeta_{2}$ follow the same Gamma distribution
$\Gamma\left(1,\sigma_{z}^{2}\right)$.

When $\text{SINR}=50\text{ dB}$, the noise variance $\sigma_{z}^{2}$
is pretty small so that we have $X\approx Y$, and the reward from
the environment could be either positive ($+1$) or negative ($-1$).
Therefore, it is fine to let $c_{u}=c_{l}=1$. Ideally, $X<1$ should
lead to a negative reward, that is, $Prob(Y>c_{u}=1)=0$, as shown
in the curve $\text{SINR}=50\text{ dB}$ in Fig. \ref{fig:The-probability-of}

In the case of $\text{SINR}=20\text{ dB}$, the noise effect on the
received power becomes serious. If we still assume that $c_{u}=1$
at $\text{SINR}=20\text{ dB}$, we can find that the probability that
the agent get a positive reward when $X<1$ is greater than 0. For
example, when $X=0.9$, the probability that the agent get a positive
reward is $Prob(Y>c_{u}=1)=0.35$. The objective of the upper and
lower thresholds are used to limit the reward from the environment
when the values of the received power are unreliable. Increasing the
value of $c_{u}$ can effectively decrease this kind of error probability.
However, it does not make sense to let $c_{u}$ be very large because
it will make the reward equal to zero even when $X>1$, which is not
beneficial for Q-learning. In the same manner, we can adjust the value
of the other threshold $c_{l}$ to reduce the probability of getting
the fail reward.

\section{Derivation of (\ref{eq: x_hat})\label{sec:Derivation-of-()}}

The objective function of the problem (\ref{eq: x_hat}) is the generalized
Rayleigh quotient \cite{Meyer2000}. To convert the problem of maximizing
SINR to a simpler one of maximizing a normalized quadratic form, we
define a vector $\tilde{\mathbf{x}}_{u}=\mathbf{B}_{u,t}^{0.5}\mathbf{x}_{u}$,
which is equivalent to $\mathbf{x}_{u}=\mathbf{B}_{u,t}^{-0.5}\tilde{\mathbf{x}}_{u}$.
Replacing $\mathbf{x}_{u}$ with $\mathbf{B}_{u,t}^{-0.5}\tilde{\mathbf{x}}_{u}$,
the objective function of the problem becomes
\begin{equation}
\frac{\tilde{\mathbf{x}}_{u}^{H}\mathbf{B}_{u,t}^{-0.5}\mathbf{A}_{u,t}\mathbf{B}_{u,t}^{-0.5}\tilde{\mathbf{x}}_{u}}{\left\Vert \tilde{\mathbf{x}}_{u}\right\Vert _{2}^{2}}.\label{eq: new_obj}
\end{equation}

To maximize (\ref{eq: new_obj}) is equivalent to maximize the numerator.
Let $\tilde{\mathbf{x}}_{u}$ be the eigenvector of $\mathbf{B}_{u,t}^{-0.5}\mathbf{A}_{u,t}\mathbf{B}_{u,t}^{-0.5}$
corresponding to the maximum eigenvalue, the maximum value of (\ref{eq: new_obj})
is therefore given by
\begin{align}
\underset{\tilde{\mathbf{x}}_{u}}{\max}\:\frac{\tilde{\mathbf{x}}_{u}^{H}\mathbf{B}_{u,t}^{-0.5}\mathbf{A}_{u,t}\mathbf{B}_{u}^{-0.5}\tilde{\mathbf{x}}_{u}}{\left\Vert \tilde{\mathbf{x}}_{u}\right\Vert _{2}^{2}} & =\lambda_{\max}(\mathbf{B}_{u,t}^{-0.5}\mathbf{A}_{u,t}\mathbf{B}_{u,t}^{-0.5}),
\end{align}
where $\lambda_{\max}(\mathbf{B}_{u,t}^{-0.5}\mathbf{A}_{u,t}\mathbf{B}_{u,t}^{-0.5})$
is the maximum eigenvalue of $\mathbf{B}_{u,t}^{-0.5}\mathbf{A}_{u,t}\mathbf{B}_{u,t}^{-0.5}$.
As a result, we have the optimal solution of $\mathbf{x}_{u}$ subject
to the constraint $\left\Vert \mathbf{x}_{u}\right\Vert _{2}=1$:
\begin{align}
\mathbf{x}_{u}^{\star} & =\frac{\mathbf{B}_{u,t}^{-0.5}\tilde{\mathbf{x}}_{u}}{\left\Vert \mathbf{B}_{u,t}^{-0.5}\tilde{\mathbf{x}}_{u}\right\Vert _{2}}\nonumber \\
 & =\frac{\mathbf{B}_{u,t}^{-0.5}\mathbf{e}_{\max}(\mathbf{B}_{u,t}^{-0.5}\mathbf{A}_{u,t}\mathbf{B}_{u,t}^{-0.5})}{\left\Vert \mathbf{B}_{u,t}^{-0.5}\mathbf{e}_{\max}(\mathbf{B}_{u,t}^{-0.5}\mathbf{A}_{u,t}\mathbf{B}_{u,t}^{-0.5})\right\Vert _{2}},
\end{align}
where $\mathbf{e}_{\max}(\mathbf{B}_{u,t}^{-0.5}\mathbf{A}_{u,t}\mathbf{B}_{u,t}^{-0.5})$
is the dominant eigenvector of $\mathbf{B}_{u,t}^{-0.5}\mathbf{A}_{u,t}\mathbf{B}_{u,t}^{-0.5}$.

\bibliographystyle{IEEEtran}
\bibliography{IEEEabrv,reference}

\begin{thebibliography}{10}
\providecommand{\url}[1]{#1}
\csname url@samestyle\endcsname
\providecommand{\newblock}{\relax}
\providecommand{\bibinfo}[2]{#2}
\providecommand{\BIBentrySTDinterwordspacing}{\spaceskip=0pt\relax}
\providecommand{\BIBentryALTinterwordstretchfactor}{4}
\providecommand{\BIBentryALTinterwordspacing}{\spaceskip=\fontdimen2\font plus
\BIBentryALTinterwordstretchfactor\fontdimen3\font minus
  \fontdimen4\font\relax}
\providecommand{\BIBforeignlanguage}[2]{{%
\expandafter\ifx\csname l@#1\endcsname\relax
\typeout{** WARNING: IEEEtran.bst: No hyphenation pattern has been}%
\typeout{** loaded for the language `#1'. Using the pattern for}%
\typeout{** the default language instead.}%
\else
\language=\csname l@#1\endcsname
\fi
#2}}
\providecommand{\BIBdecl}{\relax}
\BIBdecl

\bibitem{Nadeem2019}
\BIBentryALTinterwordspacing
Q.~Nadeem, A.~Kammoun, A.~Chaaban, M.~Debbah, and M.~Alouini, ``Asymptotic
  analysis of large intelligent surface assisted mimo communication,''
  \emph{Submitted to IEEE Trans. Wireless Commun.}, 2019. [Online]. Available:
  \url{https://arxiv.org/pdf/1903.08127.pdf}
\BIBentrySTDinterwordspacing

\bibitem{HosseinMotlagh2016}
N.~{Hossein Motlagh}, T.~{Taleb}, and O.~{Arouk}, ``Low-altitude unmanned
  aerial vehicles-based internet of things services: Comprehensive survey and
  future perspectives,'' \emph{IEEE IoT J.}, vol.~3, no.~6, pp. 899--922, Dec
  2016.

\bibitem{Uluturk2019}
I.~{Uluturk}, I.~{Uysal}, and K.~{Chen}, ``Efficient 3d placement of access
  points in an aerial wireless network,'' in \emph{IEEE Annu. Consumer Commun.
  Netw. Conf. (CCNC)}, Las Vegas, NV, USA, Jan. 2019.

\bibitem{Waharte2010}
S.~{Waharte} and N.~{Trigoni}, ``Supporting search and rescue operations with
  uavs,'' in \emph{Int. Conf. on Emerg. Security Technol.}, Canterbury, UK,
  Sep. 2010, pp. 142--147.

\bibitem{Zeng2016}
Y.~{Zeng}, R.~{Zhang}, and T.~J. {Lim}, ``Wireless communications with unmanned
  aerial vehicles: opportunities and challenges,'' \emph{IEEE Commun. Mag.},
  vol.~54, no.~5, pp. 36--42, May 2016.

\bibitem{Fiandrino2019}
C.~{Fiandrino}, H.~{Assasa}, P.~{Casari}, and J.~{Widmer}, ``Scaling
  millimeter-wave networks to dense deployments and dynamic environments,''
  \emph{Proc. IEEE}, vol. 107, no.~4, pp. 732--745, Apr. 2019.

\bibitem{Rappaport2014}
T.~Rappaport, R.~Heath, R.~Daniels, and J.~Murdock, \emph{Millimeter Wave
  Wireless Communications}.\hskip 1em plus 0.5em minus 0.4em\relax Prentice
  Hall, 2014.

\bibitem{Xiao2017a}
M.~{Xiao}, S.~{Mumtaz}, Y.~{Huang}, L.~{Dai}, Y.~{Li}, M.~{Matthaiou}, G.~K.
  {Karagiannidis}, E.~{Bj{\"o}rnson}, K.~{Yang}, C.~{I}, and A.~{Ghosh},
  ``Millimeter wave communications for future mobile networks,'' \emph{IEEE J.
  Sel. Areas Commun.}, vol.~35, no.~9, pp. 1909--1935, Sep. 2017.

\bibitem{Samimi2015}
M.~K. Samimi and T.~S. Rappaport, ``3-{D} statistical channel model for
  millimeter-wave outdoor mobile broadband communications,'' in \emph{IEEE Int.
  Conf. on Commun. (ICC)}, London, UK, Jun. 2015, pp. 2430--2436.

\bibitem{3GPP38900}
{3GPP TR 38.900 V14.3.1}, ``Study on channel model for frequency spectrum above
  6 {GHz} ({R}elease 14),'' Tech. Rep., 2017.

\bibitem{Liberti1999}
J.~Liberti and T.~Rappaport, \emph{Smart antennas for wireless communications:
  IS-95 and third generation CDMA applications}.\hskip 1em plus 0.5em minus
  0.4em\relax Prentice Hall, 1999.

\bibitem{Hajimiri2005}
A.~Hajimiri, H.~Hashemi, A.~Natarajan, X.~Guan, and A.~Komijani, ``Integrated
  phased array systems in silicon,'' \emph{Proc. IEEE}, vol.~93, no.~9, pp.
  1637--1655, Sep. 2005.

\bibitem{Bjornson2019}
E.~{Bj{\"o}rnson}, L.~{Van der Perre}, S.~{Buzzi}, and E.~G. {Larsson},
  ``Massive mimo in sub-6 {G}hz and mmwave: Physical, practical, and use-case
  differences,'' \emph{IEEE Wireless Commun.}, vol.~26, no.~2, pp. 100--108,
  Apr. 2019.

\bibitem{Zhang2005}
X.~Zhang, A.~F. Molisch, and S.-Y. Kung, ``Variable-phase-shift-based
  {RF}-baseband codesign for {MIMO} antenna selection,'' \emph{IEEE Trans.
  Signal Process.}, vol.~53, no.~11, pp. 4091--4103, Nov. 2005.

\bibitem{Ayach2014}
O.~E. Ayach, S.~Rajagopal, S.~Abu-Surra, Z.~Pi, and R.~W. Heath, ``Spatially
  sparse precoding in millimeter wave {MIMO} systems,'' \emph{IEEE Trans.
  Wireless Commun.}, vol.~13, no.~3, pp. 1499--1513, Mar. 2014.

\bibitem{Chiang2017_ICASSP}
H.~Chiang, W.~Rave, T.~Kadur, and G.~Fettweis, ``A low-complexity beamforming
  method by orthogonal codebooks for millimeter wave links,'' in \emph{IEEE
  Int. Conf. on Acoust., {S}peech and Signal Process. (ICASSP)}, New Orleans,
  LA, USA, Mar. 2017, pp. 3375 -- 3379.

\bibitem{Chiang2018_JSTSP}
------, ``Hybrid beamforming based on implicit channel state information for
  millimeter wave links,'' \emph{IEEE J. Sel. Top. Signal Process.}, vol.~12,
  no.~2, pp. 326--339, May 2018.

\bibitem{Fotouhi2019}
A.~{Fotouhi}, H.~{Qiang}, M.~{Ding}, M.~{Hassan}, L.~G. {Giordano},
  A.~{Garcia-Rodriguez}, and J.~{Yuan}, ``Survey on uav cellular
  communications: Practical aspects, standardization advancements, regulation,
  and security challenges,'' \emph{IEEE Commun. Surveys Tutorials}, 2019.

\bibitem{Zhou2019}
P.~{Zhou}, X.~{Fang}, Y.~{Fang}, R.~{He}, Y.~{Long}, and G.~{Huang}, ``Beam
  management and self-healing for mmwave uav mesh networks,'' \emph{IEEE Trans.
  Veh. Technol.}, vol.~68, no.~2, pp. 1718--1732, Feb. 2019.

\bibitem{Kadur2018}
T.~{Kadur}, W.~{Rave}, H.~{Chiang}, and G.~{Fettweis}, ``Experimental
  validation of a robust beam alignment algorithm in an indoor environment,''
  in \emph{IEEE Wireless Commun. and Networking Conf. (WCNC)}, Barcelona,
  Spain, Apr. 2018.

\bibitem{Booth2019}
M.~B. {Booth}, V.~{Suresh}, N.~{Michelusi}, and D.~J. {Love}, ``Multi-armed
  bandit beam alignment and tracking for mobile millimeter wave
  communications,'' \emph{IEEE Commun. Lett.}, vol.~23, no.~7, pp. 1244--1248,
  July 2019.

\bibitem{Va2016}
V.~{Va}, H.~{Vikalo}, and R.~W. {Heath}, ``Beam tracking for mobile millimeter
  wave communication systems,'' in \emph{IEEE Global Conf. on Signal and Inf.
  Process. (GlobalSIP)}, Washington, DC, USA, Dec 2016, pp. 743--747.

\bibitem{Supancic2017}
J.~{Supancic} and D.~{Ramanan}, ``Tracking as online decision-making: Learning
  a policy from streaming videos with reinforcement learning,'' in \emph{IEEE
  Int. Conf. on Computer Vision (ICCV)}, Venice, Italy, Oct. 2017, pp.
  322--331.

\bibitem{Klautau2018}
A.~{Klautau}, P.~{Batista}, N.~{Gonz\'{a}lez-Prelcic}, Y.~{Wang}, and R.~W.
  {Heath}, ``5{G} mimo data for machine learning: Application to beam-selection
  using deep learning,'' in \emph{Inform. Theory and Applicat. Workshop (ITA)},
  San Diego, CA, USA, Feb. 2018.

\bibitem{Chen2018}
Y.~{Chen}, W.~{Cheng}, and L.~{Wang}, ``Learning-assisted beam search for
  indoor mmwave networks,'' in \emph{IEEE Wireless Commun. and Netw. Conf.
  Workshops (WCNCW)}, Barcelona, Spain, Apr. 2018, pp. 320--325.

\bibitem{Ke2019}
Y.~{Ke}, H.~{Gao}, W.~{Xu}, L.~{Li}, L.~{Guo}, and Z.~{Feng}, ``Position
  prediction based fast beam tracking scheme for multi-user uav-mmwave
  communications,'' in \emph{IEEE Int. Conf. on Commun. (ICC)}, Shanghai,
  China, May 2019.

\bibitem{Watkins1992}
\BIBentryALTinterwordspacing
C.~J. C.~H. Watkins and P.~Dayan, ``Q-learning,'' \emph{Machine Learning},
  vol.~8, no.~3, pp. 279--292, May 1992. [Online]. Available:
  \url{https://doi.org/10.1007/BF00992698}
\BIBentrySTDinterwordspacing

\bibitem{Sutton2018}
\BIBentryALTinterwordspacing
R.~S. Sutton and A.~G. Barto, \emph{Reinforcement Learning: An Introduction},
  2nd~ed.\hskip 1em plus 0.5em minus 0.4em\relax The MIT Press, 2018. [Online].
  Available: \url{http://incompleteideas.net/book/the-book-2nd.html}
\BIBentrySTDinterwordspacing

\bibitem{Li2019}
\BIBentryALTinterwordspacing
L.~Li, H.~Ren, Q.~Cheng, K.~Xue, W.~Chen, M.~Debbah, and Z.~Han,
  ``Millimeter-wave networking in sky: A machine learning and mean field game
  approach for joint beamforming and beam-steering,'' \emph{Submitted to IEEE
  J. Sel. Areas Commun.}, 2019. [Online]. Available:
  \url{http://www.laneas.com/sites/default/files/publications/4223/JSAC-v9.2.pdf}
\BIBentrySTDinterwordspacing

\bibitem{Alkhateeb2015a}
A.~Alkhateeb, G.~Leus, and R.~W. Heath, ``Limited feedback hybrid precoding for
  multi-user millimeter wave systems,'' \emph{IEEE Trans. Wireless Commun.},
  vol.~14, no.~11, pp. 6481--6494, Nov. 2015.

\bibitem{Sohrabi2016}
F.~Sohrabi and W.~Yu, ``Hybrid digital and analog beamforming design for
  large-scale antenna arrays,'' \emph{IEEE J. Sel. Topics Signal Process.},
  vol.~10, no.~3, pp. 501--513, Apr. 2016.

\bibitem{Yin2002}
H.~Yin and H.~Liu, ``Performance of space-division multiple-access ({SDMA})
  with scheduling,'' \emph{IEEE Trans. Wireless Commun.}, vol.~1, no.~4, pp.
  611--618, Oct. 2002.

\bibitem{Adhikary2014}
A.~Adhikary, E.~A. Safadi, M.~K. Samimi, R.~Wang, G.~Caire, T.~S. Rappaport,
  and A.~F. Molisch, ``Joint spatial division and multiplexing for mm-wave
  channels,'' \emph{IEEE J. Sel. Areas Commun.}, vol.~32, no.~6, pp.
  1239--1255, Jun. 2014.

\bibitem{Balanis2005}
C.~A. Balanis, \emph{Antenna Theory: Analysis and Design}.\hskip 1em plus 0.5em
  minus 0.4em\relax Wiley-Interscience, 2005.

\bibitem{Sayeed2002}
A.~M. {Sayeed}, ``Deconstructing multiantenna fading channels,'' \emph{IEEE
  Trans. Signal Process.}, vol.~50, no.~10, pp. 2563--2579, Oct 2002.

\bibitem{Yang2019a}
L.~{Yang} and W.~{Zhang}, ``Beam tracking and optimization for uav
  communications,'' \emph{IEEE Trans. Wireless Commun. (Early Access)}, 2019.

\bibitem{Tan1993}
M.~Tan, ``Multi-agent reinforcement learning: Independent vs. cooperative
  agents,'' in \emph{Proc. 10th Intl. Conf. on Machine Learning}.\hskip 1em
  plus 0.5em minus 0.4em\relax Amherst, MA, USA: Morgan Kaufmann Publishers
  Inc., 1993, pp. 330--337.

\bibitem{Chen2019}
K.~{Chen} and H.~{Hung}, ``Wireless robotic communication for collaborative
  multi-agent systems,'' in \emph{IEEE Int. Conf. on Commun. (ICC)}, Shanghai,
  China, May 2019.

\bibitem{Meyer2000}
C.~D. Meyer, \emph{Matrix Analysis and Applied Linear Algebra}, C.~D. Meyer,
  Ed.\hskip 1em plus 0.5em minus 0.4em\relax Philadelphia, PA, USA: Society for
  Industrial and Applied Mathematics, 2000.

\end{thebibliography}

\end{document}